\documentclass[authoryear]{elsarticle}

\usepackage[a4paper,left=2.5cm,right=2.5cm,top=2cm,bottom=2cm]{geometry}
\usepackage{amsmath,amssymb}
\usepackage[unicode,colorlinks=true]{hyperref}

\newenvironment{itemize*}
{\begin{itemize}
\setlength{\itemsep}{0pt}
\setlength{\parskip}{0pt}}
{\end{itemize}}

\renewenvironment{thebibliography}[1]{
\begin{oldthebibliography}{#1}
\setlength{\itemsep}{0.8ex}
}
{\end{oldthebibliography}}

\journal{European Journal of Mechanics - A/Solids}

\begin{document}

\begin{frontmatter}

\title{Finite Element Simulation of Dense Wire Packings}

\author[a]{R. Vetter\corref{cor1}}
\ead{vetterro@ethz.ch}
\cortext[cor1]{Corresponding author.}
\author[a]{F. K. Wittel}
\author[a,b]{N. Stoop}
\author[a]{H. J. Herrmann}

\address[a]{Computational Physics for Engineering Materials, IfB, ETH Zurich, Schafmattstrasse 6, CH-8093 Zurich, Switzerland}
\address[b]{IAS Institute of Applied Simulations, ZHAW Zurich University of Applied Sciences, CH-8820 W\"adenswil, Switzerland}

\begin{abstract}
A finite element program is presented to simulate the process of packing and coiling elastic wires in two- and three-dimensional confining cavities. The wire is represented by third order beam elements and embedded into a corotational formulation to capture the geometric nonlinearity resulting from large rotations and deformations. The hyperbolic equations of motion are integrated in time using two different integration methods from the Newmark family: an implicit iterative Newton-Raphson line search solver, and an explicit predictor-corrector scheme, both with adaptive time stepping. These two approaches reveal fundamentally different suitability for the problem of strongly self-interacting bodies found in densely packed cavities. Generalizing the spherical confinement symmetry investigated in recent studies, the packing of a wire in hard ellipsoidal cavities is simulated in the frictionless elastic limit. Evidence is given that packings in oblate spheroids and scalene ellipsoids are energetically preferred to spheres.
\end{abstract}


\end{frontmatter}

\section{Introduction}
\label{sec:intro}

The dense packing and crumpling behavior of long, slender objects subject to spatial confinement is a recurring topic in various fields in natural sciences. Biophysicists that find long DNA chains densely packed in viral capsids \citep[e.g.,][]{KTBG01,GM07,RIEW07,PH08} or surgeons that treat saccular aneurysms in brain arteries by means of endovascular coiling \citep{GVDD91,GVSM91} are just two out of many examples why the coiling of thin rods inside of a confined space has attracted increasing interest in recent research. In particular the minimally invasive surgical treatment of saccular aneurysms, where a platinum wire is pushed through a small opening of the sphere-like bulge, calls for a deeper understanding of the emerging morphologies, since high packing densities favor a better long-term stability of the embolization \citep{TIAKTT02}.

In the past decade, the crumpling of thin wires has been studied in two dimensions experimentally by \cite{DGS02,DGS03,DOG06,DG07,GBCD08} followed by numerical simulations by \cite{SWH08}, from a primarily physical perspective. It wasn't until very recently that experiments were extended to three dimensions \citep{GBA08,SNWHH11}. A numerical study of the three-dimensional case by \cite{SNWHH11} has unveiled a range of interesting morphological phases largely dependent on boundary conditions and internal twist rather than the stiffness or amount of friction of the wire. Stoop et al.~used a discrete element model to represent the wire dynamics.

In this paper, a higher-order representation for elastic wires packed inside of two- or three-dimensional cavities is presented, using a finite element scheme. The finite element model is more flexible toward adaptive mesh refinement and adaptive damping, and further amenities include a less involved incorporation of plasticity effects, as well as guaranteed convergence. Using the presented model, we generalize previous studies on spherical cavity shapes to ellipsoidal ones. The results are indicating that wires coiled inside of flexible cavities (e.g.~biomaterial antra) will tend to assume non-spherical bulk shapes, differing to some degree from the emerging morphologies recently found in perfectly spherical confinement.

The paper is organized as follows: Section \ref{sec:fem} presents the construction of a third order beam finite element model representing an isotropic wire that is pushed into (or pulled out of) a hard ellipsoidal cavity. Both theoretical and technical aspects of the geometrical nonlinearity arising from large rotations and deformations are addressed and amended by the incorporation of wire-cavity contact and wire-wire contact. The main components of the strain energy are expressed in terms of the finite wire elements. Section \ref{sec:integration} addresses the integration of the hyperbolic equations of motion in time, using an implicit and an explicit scheme of the Newmark family \citep{N59}. The presented beam theory is verified by comparison to literature in Section \ref{sec:verification}, where we also argue that a purely implicit treatment of wire-wire contacts within the finite element framework is unfeasible in the context of dense packings. Section \ref{sec:results} briefly presents new simulation results for a selection of different ellipsoidal isochoric and isoareal cavity shapes, unveiling their respective energy levels. In Section \ref{sec:conclusions}, the results are summarized and an outlook to future extensions of the present work is given.

\section{Finite element model}
\label{sec:fem}

\subsection{Equations of motion}

The wire is represented by a one-dimensional regular mesh with $N$ elements $\{e\}$, i.e.~$N+1$ nodes $\{n\}$, with uniform spacing $h$. The undeformed mesh is positioned at ${\bf x}\in\mathbb{R}^{3(N+1)}$, aligned to the global $x$-axis in the simulations. Each mesh node carries six degrees of freedom (DOFs): three Cartesian displacements $u_n$, $v_n$, $w_n$ in $x$-, $y$-, and $z$-direction, respectively, and three Euler angles $\varphi_n$, $\theta_n$, $\psi_n$ for rotations around these respective axes, giving each node an orientation. The center line of the wire is interpolated along each element using cubic Hermite shape functions. The motion of the wire is described by the nonlinear system of hyperbolic differential equations
\begin{equation}
\label{eq:full_newton}
{\bf M}\ddot{{\bf u}} + {\bf C}\dot{{\bf u}} + {\bf f}_{\textrm{int}}({\bf u}) = {\bf f}_{\textrm{ext}}({\bf u}).
\end{equation}
Here, ${\bf M}$ is the wire's mass matrix, ${\bf C}$ is a damping matrix and ${\bf f}_{\textrm{int}}({\bf u})$ represents the solution-dependent internal elastic forces. Each of the matrices is an element of $\mathbb{R}^{6(N+1)\times 6(N+1)}$. The global solution vector ${\bf u}={\bf u}(t)\in\mathbb{R}^{6(N+1)}$ consists of the six DOFs for each node at time $t$. All external forces and torques acting on the wire at time $t$ enter the solution-dependent right-hand side vector ${\bf f}_{\textrm{ext}}({\bf u})$. This includes self-contact forces between wire elements and forces due to repulsion from the cavity. The conventional DOF ordering ${\bf u} = [{\bf u}_0^{\mathrm{T}}, {\bf u}_1^{\mathrm{T}}, ..., {\bf u}_N^{\mathrm{T}}]^{\mathrm{T}}$ is adopted in the following, with nodal contributions
\begin{equation}
\label{eq:un}
{\bf u}_n = [u_n,v_n,w_n,\varphi_n,\theta_n,\psi_n]^{\mathrm{T}}\in\mathbb{R}^6,\quad n = 0,1,...,N.
\end{equation}
To calculate local element contributions, as is common in the finite element method, we employ the local element ordering
\begin{equation}
\label{eq:ue}
{\bf u}_e = [u_1,v_1,w_1,\varphi_1,\theta_1,\psi_1,u_2,v_2,w_2,\varphi_2,\theta_2,\psi_2]^{\mathrm{T}}\in\mathbb{R}^{12},
\end{equation}
where the subscript 1 refers to the element's left node and 2 its right counterpart. In order to avoid the singularities occurring in Eulerian representations of large rotations, auxiliary nodal unit quaternions ${\bf q}_n$ are used in addition. They are initialized according to
\begin{equation}
\label{eq:qvec}
{\bf q}_n = q_0 + i\,\overline{\bf q} = \begin{bmatrix}q_0\\q_1\\q_2\\q_3\end{bmatrix} = \begin{bmatrix}
\cos(\varphi_n/2)\cos(\theta_n/2)\cos(\psi_n/2) + \sin(\varphi_n/2)\sin(\theta_n/2)\sin(\psi_n/2)\\
\sin(\varphi_n/2)\cos(\theta_n/2)\cos(\psi_n/2) - \cos(\varphi_n/2)\sin(\theta_n/2)\sin(\psi_n/2)\\
\cos(\varphi_n/2)\sin(\theta_n/2)\cos(\psi_n/2) + \sin(\varphi_n/2)\cos(\theta_n/2)\sin(\psi_n/2)\\
\cos(\varphi_n/2)\cos(\theta_n/2)\sin(\psi_n/2) - \sin(\varphi_n/2)\sin(\theta_n/2)\cos(\psi_n/2)
\end{bmatrix}.
\end{equation}
Each of them defines a nodal unit triad
\begin{equation}
\label{eq:triad}
{\bf T}_n = [{\bf t}_1,{\bf t}_2,{\bf t}_3] = \begin{bmatrix}
q_0^2 + q_1^2 - q_2^2 - q_3^2 & 2(q_1 q_2 - q_0 q_3) & 2(q_0 q_2 + q_1 q_3)\\
2(q_1 q_2 + q_0 q_3) & q_0^2 - q_1^2 + q_2^2 - q_3^2 & 2(q_2 q_3 - q_0 q_1)\\
2(q_1 q_3 - q_0 q_2) & 2(q_0 q_1 + q_2 q_3) & q_0^2 - q_1^2 - q_2^2 + q_3^2
\end{bmatrix},
\end{equation}
storing the orientation of each mesh node $n$ in space. Given a small increment ${\bf\Delta u}_n$ to (\ref{eq:un}), be it obtained in whatever way, the quaternion can be updated by applying the non-commutative Hamilton product
\begin{equation}
\label{eq:quat_prod}
{\bf q}_n := {\bf \Delta q}_n\,{\bf q}_n = \Delta q_0\,q_0 - ({\bf\Delta\overline q})^{\textrm{T}}{\bf\overline q} + i\,(q_0\,{\bf\Delta\overline q} + \Delta q_0\,{\bf\overline q} + {\bf\Delta\overline q}\times{\bf\overline q}),
\end{equation}
where ${\bf \Delta q}_n$ is built using Eq.~(\ref{eq:qvec}) on the angular increments $\Delta\varphi_n,\Delta\theta_n,\Delta\psi_n$ from ${\bf\Delta u}_n$. The purpose in defining triads as in Eq.~(\ref{eq:triad}) is elucidated in Section \ref{sec:cr}, where the nodes may be arbitrarily oriented to allow for very large deformations.

For simplicity, the mass matrix ${\bf M}$ is diagonalized by means of \textit{direct mass lumping} \citep{TPB71}:
\begin{equation}
{\bf M}_n = \textrm{diag}(m_n,m_n,m_n,J_n,J_n,J_n)
\end{equation}
with nodal masses $m_n = Ah\rho$, where $A=\pi r^2$ is the cross-section area of the tubular wire and $\rho$ denotes its mass density. For the nodal moments of inertia $J_n$, the point masses $m_n$ are uniformly distributed within balls of the same radius as the wire, resulting in $J_n = \frac{2}{5}m_nr^2$. For damping, one may use the Rayleigh ansatz ${\bf C} = \alpha{\bf M} + \beta{\bf K}$, $\alpha,\beta\ge 0$. In the present simulations, this is simplified to ${\bf C}=c\,{\bf I}_{6(N+1)}$, where $c\ge 0$ is a scalar viscous damping coefficient acting equally on all translational and angular velocities, and ${\bf I}_k$ denotes the $k\!\times\! k$ identity matrix.

\subsection{Three-dimensional beam theory}

The bending deformation of the tubular wire is modeled by Reddy's simplified third-order beam theory (RBT) \citep{RWL97,R97} that includes a quadratic transverse shear stress distribution, contrary to previously published discrete element simulations \citep{SWH08,SNWHH11}. RBT completely alleviates the problem of shear locking \citep{TD81,PB82} as it is known in Timoshenko beam theory \citep{T21,T22}, without complicating the stiffness matrix. Assuming independent bending in both the $xy$ and $xz$ planes, in particular vanishing mixed-bending Cauchy stresses $\sigma_{yz} = \sigma_{zy} = 0$, the two-dimensional linear RBT element stiffness matrix for constant beam material and geometry can be extended to 3D as follows:
\begin{equation}
\label{eq:reddy_elem_stiff_3d}
{\bf K}_{e,\textrm{lin}} = \begin{bmatrix}
k_{u} & 0 & 0 & 0 & 0 & 0 & -k_{u} & 0 & 0 & 0 & 0 & 0\\
& k_{v} & 0 & 0 & 0 & k_{v\psi} & 0 & -k_{v} & 0 & 0 & 0 & k_{v\psi}\\
& & k_{w} & 0 & -k_{w\theta} & 0 & 0 & 0 & -k_{w} & 0 & -k_{w\theta} & 0\\
& & & k_{\varphi} & 0 & 0 & 0 & 0 & 0 & -k_{\varphi} & 0 & 0\\
& & & & k_{\theta} & 0 & 0 & 0 & k_{w\theta} & 0 & k_{\theta\theta} & 0\\
& & & & & k_{\psi} & 0 & -k_{v\psi} & 0 & 0 & 0 & k_{\psi\psi}\\
& & & & & & k_{u} & 0 & 0 & 0 & 0 & 0\\
& & & & & & & k_{v} & 0 & 0 & 0 & -k_{v\psi}\\
& & \multicolumn{2}{c}{\textrm{symm.}} & & & & & k_{w} & 0 & k_{w\theta} & 0\\
& & & & & & & & & k_{\varphi} & 0 & 0\\
& & & & & & & & & & k_{\theta} & 0\\
& & & & & & & & & & & k_{\psi}\\
\end{bmatrix}
\end{equation}
with entries
\begin{equation}
\label{eq:stiff_entries}
k_w = \frac{12EI_{yy}}{\mu_{xz} h^3},\qquad k_{\theta} = \frac{4EI_{yy}}{\mu_{xz} h}\lambda_{xz},\qquad k_{w\theta} = \frac{6EI_{yy}}{\mu_{xz} h^2},\qquad k_{\theta\theta} = \frac{2EI_{yy}}{\mu_{xz} h}\xi_{xz},
\end{equation}
\begin{equation}
\label{eq:stiff_entries_z}
k_v = \frac{12EI_{zz}}{\mu_{xy} h^3},\qquad k_{\psi} = \frac{4EI_{zz}}{\mu_{xy} h}\lambda_{xy},\qquad k_{v\psi} = \frac{6EI_{zz}}{\mu_{xy} h^2},\qquad k_{\psi\psi} = \frac{2EI_{zz}}{\mu_{xy} h}\xi_{xy},
\end{equation}
\begin{equation}
\label{eq:greeks}
\mu_{xz} = 1+12\Omega_{xz},\qquad\lambda_{xz}=1+3\Omega_{xz},\qquad\xi_{xz}=1-6\Omega_{xz}.
\end{equation}
$\Omega_{xz}$ is the parameter determining the order of the theory:
\begin{equation}
\Omega_{xz}=\begin{cases}0 & \textrm{for Euler-Bernoulli theory (EBT)}\\ \frac{\hat{D}_{xx}}{\hat{A}_{xz}h^2} & \textrm{for RBT}\end{cases}.
\end{equation}
In RBT, the stiffness coefficients
\begin{equation}
\begin{array}{r@{}l@{}l@{}lr@{}l@{}l@{}l}
\hat{D}_{xx} &\;= &\overline{D}_{xx}   &\;-\;c_1 \overline{F}_{xx} , &\;\hat{A}_{xz} &\;= &\overline{A}_{xz}  &\;-\;c_2 \overline{D}_{xz},\\
\overline{D}_{xx} &\;= E\big(&I_{yy}^{(2)} &\;-\;c_1 I_{yy}^{(4)}\big), &\;\overline{A}_{xz} &\;= G\big(&A &\;-\;c_2 I_{yy}^{(2)}\big),\\
\overline{F}_{xx} &\;= E\big(&I_{yy}^{(4)} &\;-\;c_1 I_{yy}^{(6)}\big), &\;\overline{D}_{xz} &\;= G\big(&I_{yy}^{(2)} &\;-\;c_2 I_{yy}^{(4)}\big),
\end{array}
\label{eq:rbt_coeff}
\end{equation}
\begin{equation}
3c_1 = c_2 = 1/r_z^2,
\end{equation}
are needed, in which $r_z$ is the half thickness of the beam in $z$ direction. $I_{yy}^{(i)}$ is the $i$-th moment of inertia of the wire about the $y$-axis, and is calculated as
\begin{equation}
I_{yy}^{(i)} = \int_Az^i\;\textrm{d}A.
\label{eq:rbt_Iyy}
\end{equation}
Relations (\ref{eq:greeks}\textendash\ref{eq:rbt_Iyy}) hold analogously with $y$ and $z$ interchanged. $E$ denotes the Young's modulus of the isotropic wire, $G = E/2(1+\nu)$ the elastic shear modulus, and $\nu$ the Poisson ratio. The longitudinal and torsional spring constants are given by
\begin{equation}
\label{eq:stiff_entries_tens}
k_{u} = \frac{EA}{h},\qquad k_{\varphi} = \frac{GJ}{h},
\end{equation}
where the polar moment of inertia reads $J = I_{yy} + I_{zz}$ by the perpendicular axis theorem. For a wire with circular cross-section and radius $r=r_y=r_z$, one can write $I_{yy}^{(i)} = I_{zz}^{(i)} =: I^{(i)}$, which implies that
\begin{equation}
k_v = k_w,\quad k_{\psi} = k_{\theta},\quad k_{v\psi}=k_{w\theta},\quad k_{\theta\theta} = k_{\psi\psi},
\end{equation}
\begin{equation}
\label{eq:I}
I:=I^{(2)}=\frac{\pi}{4}r^4,\quad I^{(4)}=\frac{\pi}{8}r^6,\quad I^{(6)}=\frac{5\pi}{64}r^8,\quad J=2I=\frac{\pi}{2}r^4,
\end{equation}
such that $\Omega_{xy}=\Omega_{xz}=\frac{101}{180}(1+\nu)\left(\frac{r}{h}\right)^2$ in RBT. Note that in Eq.~(\ref{eq:reddy_elem_stiff_3d}) the rotational DOFs in the two bending directions have opposite sign convention such that both ``slopes'' $\theta$ and $\phi$ are equally oriented as their respective conventional Euler angles. The longitudinal displacement $u$ and the twist angle $\varphi$ are modeled with Hooke's law.

\subsection{Corotational formulation}
\label{sec:cr}

In order to comprise large rotations and deflections in a finite element simulation, geometric nonlinearity must be introduced to the theory, which can be expressed by the fact that the dependence of the internal forces ${\bf f}_{\textrm{int}}$ on ${\bf u}$ is not a mere linear product with a constant stiffness matrix ${\bf K}$. To this end, three prominent approaches have been developed, refined, mixed, and broadly applied over decades. The \textit{total Lagrangian} formulation (TL) expresses the sought solution with respect to an initial configuration. The \textit{updated Lagrangian} formulation (UL) can be shown to be effectively equivalent to TL \citep{S08}, but makes use of a different representation, where the nodal displacements and rotations are described incrementally. The \textit{corotational} formulation (CR) introduces a local reference frame for each element that continuously moves and rotates with the element, within which the theory can be considered linear. CR has been acclaimed superiority over TL and UL in accuracy \citep{MBS86,UR05}, performance \citep{BH73,RB86}, simplicity \citep{TC98} and generality \citep{RB86,C90}, and it is utilized in the following.

The corotational formulation is rooted in in the works of \cite{W69}, \cite{BH73}, and \cite{C73a,C73b}. Assuming large displacements and rotations but small strains, the idea is to corotate a dedicated coordinate frame with each beam element, exactly passing through the nodes. The 12 global DOFs per element are reduced by its five rigid body DOFs to seven local deformation DOFs when expressed with respect to this corotated frame. \cite{C90,C97} was the first to provide a beam-theory-independent, three-dimensional CR formulation with consistent treatment of the tangent stiffness matrix required by the Newton-Raphson method. Details of the algorithm can be found there, here, only the application to RBT wire elements is briefly discussed.

\begin{figure*}[h]
	\begin{center}
		\includegraphics{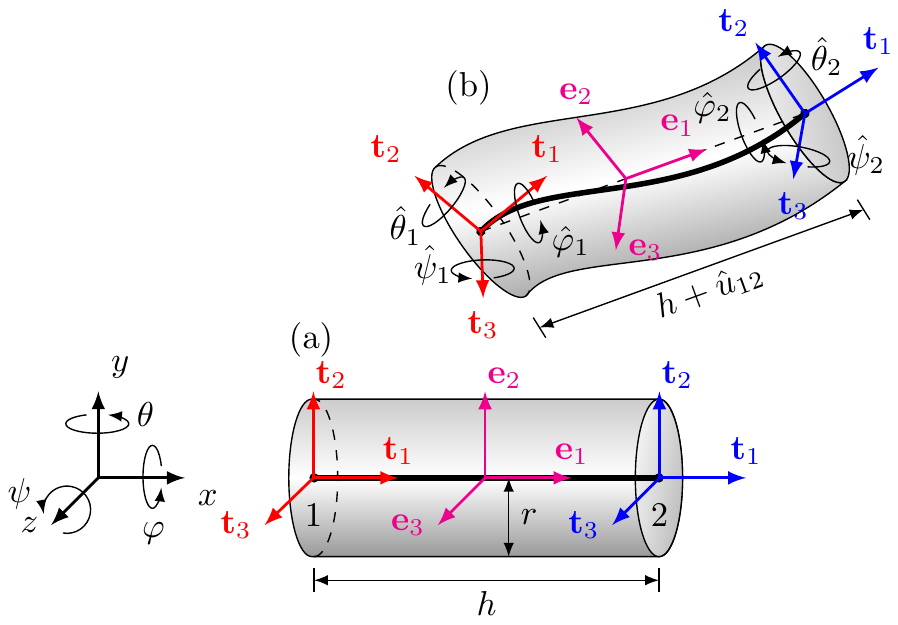}
	  \caption{Visualization of the corotated wire element: In the initial straight configuration (a), each element carries canonical triads ${\bf E}_e={\bf T}_1={\bf T}_2={\bf I}_3$. After deformation (b), the seven local element DOFs (bearing a hat) are found with the help of the triads ${\bf T}_1$ (red) and ${\bf T}_2$ (blue) storing the nodal orientations, and the corotated element frame ${\bf E}_e$ (magenta) encoding the element orientation.}
	  \label{fig:cr}
	\end{center}
\end{figure*}

Given the global nodal positions ${\bf p}_n={\bf x}_n+[u_n,v_n,w_n]^{\mathrm{T}}$ and orientations ${\bf T}_n$, each beam element $e$ is assigned a unit triad ${\bf E}_e=[{\bf e}_1,{\bf e}_2,{\bf e}_3]$ in the CR setup, called the \textit{corotated frame}, in such a way that it represents the orientation of $e$. In particular, ${\bf e}_1$ connects the two nodes:
\begin{equation}
{\bf e}_1 = \frac{{\bf p}_2-{\bf p}_1}{|{\bf p}_2-{\bf p}_1|}.
\end{equation}
The global element DOFs ${\bf u}_e$ are then expressed with respect to ${\bf E}_e$. In the following, all variables carrying a hat refer to this local element frame. By construction, we have $\hat{v}_1 = \hat{v}_2 = \hat{w}_1 = \hat{w}_2 = 0$. Additionally, the longitudinal displacement variables $u_1$, $u_2$ can be reduced to a single longitudinal spring-like deformation $\hat{u}_{12} = |{\bf p}_2-{\bf p}_1| - h$\label{eq:cr_ue}. It turns out that if ${\bf e}_2,{\bf e}_3$ are constructed from the nodal quaternions according to Crisfield, the remaining six local rotational DOFs with respect to ${\bf E}_e$ can be computed as
\begin{equation}
\label{eq:cr_local_rot}
\begin{array}{r@{}l@{}r@{}}
\hat{\varphi}_n &= \sin^{-1}\left(\frac{1}{2}({\bf t}_2^{\mathrm{T}}{\bf e}_3 - {\bf t}_3^{\mathrm{T}}{\bf e}_2)\right) - \hat{\varphi}_n^0,\\[6pt]
\hat{\theta}_n &= \sin^{-1}\left(\frac{1}{2}({\bf e}_2^{\mathrm{T}}{\bf t}_1 - {\bf t}_2^{\mathrm{T}}{\bf e}_1)\right) - \hat{\theta}_n^0,\\[6pt]
\hat{\psi}_n &= \sin^{-1}\left(\frac{1}{2}({\bf e}_3^{\mathrm{T}}{\bf t}_1 - {\bf t}_3^{\mathrm{T}}{\bf e}_1)\right) - \hat{\psi}_n^0,
\end{array}
\end{equation}
for both nodes $n$ adjacent to element $e$, each using their respective triad ${\bf T}_n = [{\bf t}_1,{\bf t}_2,{\bf t}_3] $. $\hat{\varphi}^0,\hat{\theta}^0,\hat{\psi}^0$ allow to supply the wire with intrinsic curvature or twist, i.e.~curvature or twist occurring without external loads or moments. For instance, a uniform intrinsic curvature $\kappa$ in the $xz$ plane is achieved by setting $-\hat{\theta}_1^0=\hat{\theta}_2^0=h\kappa/2$ in each element frame. This defines the seven local deformation DOFs $\hat{{\bf u}}_e = [\hat{\varphi}_1,\hat{\theta}_1,\hat{\psi}_1,\hat{u}_{12},\hat{\varphi}_2,\hat{\theta}_2,\hat{\psi}_2]^{\mathrm{T}}$ from which the rigid body motion has been completely detached, and which are shown together with the triads in Fig.~\ref{fig:cr}. The corresponding $7\!\times\! 7$ local element stiffness matrix acting on them is easily devised from Eq.~(\ref{eq:reddy_elem_stiff_3d}) using the definition of the corotated frame:
\begin{equation}
\label{eq:reddy_elem_stiff_cr}
\hat{{\bf K}}_e = \begin{bmatrix}
k_{\varphi} & 0 & 0 & 0 & -k_{\varphi} & 0 & 0\\
& k_{\theta} & 0 & 0 & 0 & k_{\theta\theta} & 0\\
& & k_{\psi} & 0 & 0 & 0 & k_{\psi\psi}\\
& & & k_{u} & 0 & 0 & 0\\
& & & & k_{\varphi} & 0 & 0\\
& \multicolumn{2}{c}{\textrm{symm.}} & & & k_{\theta} & 0\\
& & & & & & k_{\psi}\\
\end{bmatrix}.
\end{equation}
While the local element stiffness matrix remains constant throughout the simulation, nonlinearity enters the equations of motion when $\hat{{\bf K}}_e$ is transformed into the desired $12\!\times\!12$ ``nonlinear'' global element stiffness matrix ${\bf K}_e({\bf u}_e)$ via a solution-dependent global-to-local transformation matrix ${\bf F}_e({\bf u}_e)\in\mathbb{R}^{7\times 12}$:
\begin{equation}
\label{eq:cr_stiff}
{\bf K}_e({\bf u}_e) = {\bf F}_e^{\mathrm{T}}({\bf u}_e)\hat{{\bf K}}_e{\bf F}_e({\bf u}_e).
\end{equation}
Equivalently, one can write
\begin{equation}
\label{eq:cr_fint}
{\bf f}_{\textrm{int},e} = {\bf F}_e^{\mathrm{T}}({\bf u}_e)\hat{{\bf K}}_e\hat{{\bf u}}_e
\end{equation}
for the internal force vector. Note that $\hat{{\bf u}}_e$ contains the intrinsic angles $\hat{\varphi}^0,\hat{\theta}^0,\hat{\psi}^0$ via Eq.~(\ref{eq:reddy_elem_stiff_cr}), such that ${\bf f}_{\textrm{int},e}$ may be non-zero even if the element solution in global coordinates, ${\bf u}_e$, vanishes. The construction of ${\bf F}_e$ is done row-wise and comprises somewhat lengthy matrix calculus involving the three triads.

The most significant part of Crisfield's work is the derivation of a tangent stiffness matrix
\begin{equation}
\label{eq:cr_tan_stiff}
{\bf K}_{\textrm{t},e}({\bf u}_e) = \frac{\partial}{\partial{\bf u}_e}{\bf K}_e({\bf u}_e){\bf u}_e = {\bf K}_e({\bf u}_e) + {\bf K}_{\sigma,e}
\end{equation}
that is consistent with the rest of the formalism, which is needed for implicit integration of the equations of motion in time with the Newton-Raphson method. The \textit{geometric stiffness matrix} ${\bf K}_{\sigma,e}$ is an intricate sum of submatrices expressed in terms of the triads, the internal forces (\ref{eq:cr_fint}), the local DOFs $\hat{{\bf u}}_e$ and the corotational transformation matrix ${\bf F}_e$. Details are readily available in \cite{C90,C97}.

This setup allows to numerically integrate the solution vector ${\bf u}$ in time. In each time step, and in each iteration of the nonlinear solver routine, for each element, the current corotated frame ${\bf E}_e$, the seven local DOFs $\hat{{\bf u}}_e$, and the transformation matrix ${\bf F}_e$ are computed. The internal force (\ref{eq:cr_fint}) is then calculated, and if required by the integration method, the tangent stiffness matrix is built using Eqs.~(\ref{eq:cr_stiff}) and (\ref{eq:cr_tan_stiff}). After a new solution vector has been found with the applied integration scheme, the nodal quaternions are updated by applying the quaternion product (\ref{eq:quat_prod}), where the ${\bf \Delta q}_n$ are built using Eq.~(\ref{eq:qvec}) on the vectorial difference between new and old solution. The quaternions need to be renormalized on a regular basis.

\subsection{Strain energy}

For Euler-Bernoulli theory, it can easily be shown by integrating the cubic Hermite splines over the corotated elements, that the potential energy due to bending can be expressed in terms of the local DOFs by
\begin{equation}
\label{eq:ubend}
U_{\textrm{b}} = \frac{2EI}{h}\sum_e\left(\hat{\theta}_1^2 + \hat{\theta}_1\hat{\theta}_2 + \hat{\theta}_2^2 + \hat{\psi}_1^2 + \hat{\psi}_1\hat{\psi}_2 + \hat{\psi}_2^2\right),
\end{equation}
in which subscript 1 and 2 again refer to the left and right node of element $e$. In RBT, however, the Hermitian shape functions are modified to account for the shear strain approximation, and Eq.~(\ref{eq:ubend}) is valid for relatively thin wires only. In the packing simulations presented in Section \ref{sec:results}, bending shear effects turn out to be small enough to use Eq.~(\ref{eq:ubend}) as a very close approximation. Independent of the employed bending theory, the axial elastic strain energy of the wire is straightforwardly obtained from Hooke's law in the corotated frames:
\begin{equation}
U_{\textrm{s}} = \frac{EA}{2h}\sum_e\hat{u}_{12}^2.
\end{equation}
Analogously, the torsional potential energy reads
\begin{equation}
\label{eq:Ut}
U_{\textrm{t}} = \frac{GJ}{2h}\sum_e\left(\hat{\varphi}_2 - \hat{\varphi}_1\right)^2.
\end{equation}

\subsection{Wire-cavity contact}

Spatial confinement is introduced by a rigid ellipsoidal container centered at the origin, with a single small opening through which the wire is pushed inside. Hertzian contact mechanics is used to model the repulsive forces exerted on the wire nodes by the cavity. In the event that the sphere with radius $r$ around a node $n$ indents the cavity wall by a depth $D>0$, the normal force $f_{\perp}=E^*hD\pi/4$ \citep{J85} is applied to the three translational DOFs of the penetrating node, i.e.~the nodal contribution to ${\bf f}_{\textrm{ext}}$ is
\begin{equation}
\label{eq:fcav}
{\bf f}_{\textrm{cav},n} = -f_{\perp}\,{\bf n}_n\in\mathbb{R}^3.
\end{equation}
Here, ${\bf n}_n$ is the outer normal vector of the cavity surface at the contact point, and $E^*$ mixes the material parameters $(E,\nu)$ of the rod with those of the cavity, $(E_{\textrm{cav}},\nu_{\textrm{cav}})$, according to
\begin{equation}
\frac{1}{E^*} = \frac{1-\nu^2}{E} + \frac{1-\nu_{\textrm{cav}}^2}{E_{\textrm{cav}}}.
\end{equation}
The normal force is that of a Hertzian contact between two parallel cylinders, which serves as a reasonable rough estimate for the real force exerted by a large, hard cavity to a cylinder. Alternative models such as the Hertzian force between a sphere and a half-space, where the force is not linearly proportional to $D$, may be used instead if desired. The exact kind of the applied force law turns out to be of very minor importance for the present packing model.

The calculation of the shortest distance between an exact ellipsoid and a point in space is known to be equivalent to the problem of finding the roots of a sixth order polynomial \citep{H94}, for which no analytical solution is known. Thus, the indentation depth $D$ can only be found numerically, and so can the Jacobian, which needs to be specified in the Newton-Raphson scheme for truly implicit integration in time. Particularly concerning the Jacobian, this is very inconvenient. Instead, a closed-form approximation is developed here. Consider an ellipsoidal cavity given by
\begin{equation}
\label{eq:ellipsoid}
E(x,y,z) = \left(\frac{x}{R_x}\right)^2 + \left(\frac{y}{R_y}\right)^2 + \left(\frac{z}{R_z}\right)^2 = 1.
\end{equation}
For reasonably shaped ellipsoids not too far away from a sphere, i.e.~for $R_x\approx R_y\approx R_z\gg r$, a cavity contact is assumed if $\Delta\ge 0$ in the implicit formula of an \textit{effective} ellipsoid
\begin{equation}
E_{\textrm{eff}}({\bf p}_n) := \left(\frac{p_x}{R_x-r}\right)^2 + \left(\frac{p_y}{R_y-r}\right)^2 + \left(\frac{p_z}{R_z-r}\right)^2 = 1+\Delta,
\end{equation}
given the nodal position ${\bf p}_n=[p_x,p_y,p_z]^{\mathrm{T}}$. Since the Hertzian force outgrows the internal forces of the wire even for small indentations $D$ when compared to the length scales of the wire, its \textit{magnitude} is far less important than its \textit{direction}. Therefore, $D$ can legitimately be approximated using
\begin{equation}
(\overline{R}-r+D)^2 \approx p_x^2 + p_y^2 + p_z^2 \approx (\overline{R}-r)^2(1+\Delta),
\end{equation}
where $\overline{R}=(R_x+R_y+R_z)/3$ averages the ellipsoidal radii. Taking the square root yields the sought approximation
\begin{equation}
\label{eq:d_ell}
D \approx (\overline{R}-r)\big(\sqrt{1+\Delta}-1\big).
\end{equation}
A consistent approximation of the outer surface normal vector assuming that $\Delta\ll 1$ is found by the normalized gradient of the effective ellipsoid:
\begin{equation}
\label{eq:n_ell}
{\bf n}_n \approx \frac{\nabla E_{\textrm{eff}}({\bf p}_n)}{|\nabla E_{\textrm{eff}}({\bf p}_n)|} =: \frac{{\bf \tilde{p}}_n}{|{\bf \tilde{p}}_n|}.
\end{equation}
The effective direction ${\bf \tilde{p}}_n$ can be compactly written as ${\bf \tilde{p}}_n = 2\,{\bf R}^{-2}{\bf p}_n$ with a diagonal matrix
\begin{equation}
{\bf R} = \textrm{diag}(R_x-r,R_y-r,R_z-r)\in\mathbb{R}^{3\times 3}.
\end{equation}
The Jacobian ${\bf J}_{\textrm{cav},n}\in\mathbb{R}^{3\times 3}$\label{eq:Jmat} of the nodal force vector is easily assembled using the standard rules of differential calculus:
\begin{align}
\label{eq:jac_cav}
{\bf J}_{\textrm{cav},n} &= \frac{\partial{\bf f}_{\textrm{cav},n}}{\partial{\bf p}_n} = -f_{\perp}\left(\frac{\partial{\bf n}_n}{\partial{\bf p}_n}+\frac{{\bf n}_n}{D}\frac{\partial D}{\partial{\bf p}_n}\right),\\
\label{eq:dndp_ellipsoid}
\frac{\partial{\bf n}_n}{\partial{\bf p}_n} &\approx \frac{2}{|{\bf \tilde{p}}_n|}\left({\bf I}_3-\frac{1}{|{\bf \tilde{p}}_n|^2}{\bf \tilde{p}}_n{\bf \tilde{p}}_n^{\mathrm{T}}\right){\bf R}^{-2},\\
\label{eq:dDdp_ellipsoid}
\frac{\partial D}{\partial{\bf p}_n} &\approx \frac{\overline{R}-r}{2\sqrt{1+\Delta}}{\bf \tilde{p}}_n^{\mathrm{T}}.
\end{align}

The impact of the approximations in Eqs.~(\ref{eq:d_ell}) and (\ref{eq:n_ell}) is twofold. First, the magnitude of the Hertzian contact force is subject to an error that grows with the disparity of $R_x,R_y,R_z$. Since the Hertzian contact model used here is itself only a crude approximation, this has no negative consequence in practice. The only purpose is to keep the wire inside the confining container. Second, the effective ellipsoid has slightly deformed aspect ratios compared to the aimed cavity. This defect is also minor for $R_x\approx R_y\approx R_z$ and vanishes for arbitrary $R_x,R_y,R_z$ in the slender wire limit $r\to 0$. All approximations above become exact under spherical symmetry ($R_x=R_y=R_z$).\\

The axial symmetry of the straight wire is broken by imposing a small initial random deflection perpendicular to the wire on two nodes. The mesh is dynamically extended by additional elements as the wire enters the cavity. Boundary conditions are imposed by constraining nodes outside the cavity to the $x$-axis, penalizing rotations in the outermost node (no dissipation of torsional energy out of the cavity), and by applying a constant insertion velocity $\dot{u}_N\equiv v_{\textrm{in}}$. The packing density is computed using
\begin{equation}
\phi=\frac{2rL}{\pi R_x R_z}\quad\textrm{in 2D},\qquad\phi=\frac{\pi r^2L}{4/3\,\pi R_x R_y R_z}\quad\textrm{in 3D},\qquad L=\sum_e (\hat{u}_{12}+h).
\end{equation}

\subsection{Wire-wire contact}

Self-interaction of the wire happens among pairs of wire elements ($e_1,e_2)$, and the resulting forces act on the translational DOFs of the four involved nodes $n_1,n_2$ of element $e_1$ and $n_3,n_4$ of element $e_2$. For simplicity, all wire elements are treated as if they were sphero-cylindrical in shape. Since the number $N_{\textrm{c}}$ of element pairs in contact grows like the square of the packing density in the uncorrelated thin rod limit in 3D \citep{P96} (see also inset of Fig.~\ref{fig:ubend}b), an efficient treatment of these pairs is crucial for performance. Calculating the shortest distance between any two elements would lead to $\mathcal{O}(N^2)$ distance calculations, but only a small subset of them will actually be in contact. We therefore separate the problem into two steps.

First, possible contact pairs are found efficiently by using linked cell lists \citep{QB73}. The linked cell algorithm assigns every element to a cubic cell, such that only elements assigned to the same or adjacent cells have to be considered as possible contact candidates, reducing the complexity to $\mathcal{O}(N)$. The minimum cell size guaranteeing that any contact between two wire elements happens within neighboring cells is $l=2(h+r)$. To avoid having to rebuild the linked cells after each iteration step, a small additional margin can be added to the cell size, and the cells are rebuilt only after the wire has moved farther than half of that margin.

In the second step, the shortest distance between every contact candidate is calculated using the quadric algorithm by \cite{S01}, which provides an efficient implementation based on a note by \cite{E99,E00}, with the modification that the center line segments are normalized to unit length to fix the faulty detection of short parallel segments in Sunday's algorithm. The algorithm finds $s_1,s_2\in[0,1]$ such that the closest points of approach on the wire center lines read
\begin{equation}
{\bf c}_1 = {\bf p}_{n_1} + s_1({\bf p}_{n_2}-{\bf p}_{n_1}),\qquad{\bf c}_2 = {\bf p}_{n_3} + s_2({\bf p}_{n_4}-{\bf p}_{n_3}).
\end{equation}
The mutual indentation of the elements is then given by its depth $D = 2r-|{\bf \Delta c}|$ and direction ${\bf n} = {\bf \Delta c}/|{\bf \Delta c}|$, where ${\bf \Delta c}={\bf c}_1-{\bf c}_2$.

\begin{figure*}[h]
	\begin{center}
		\includegraphics{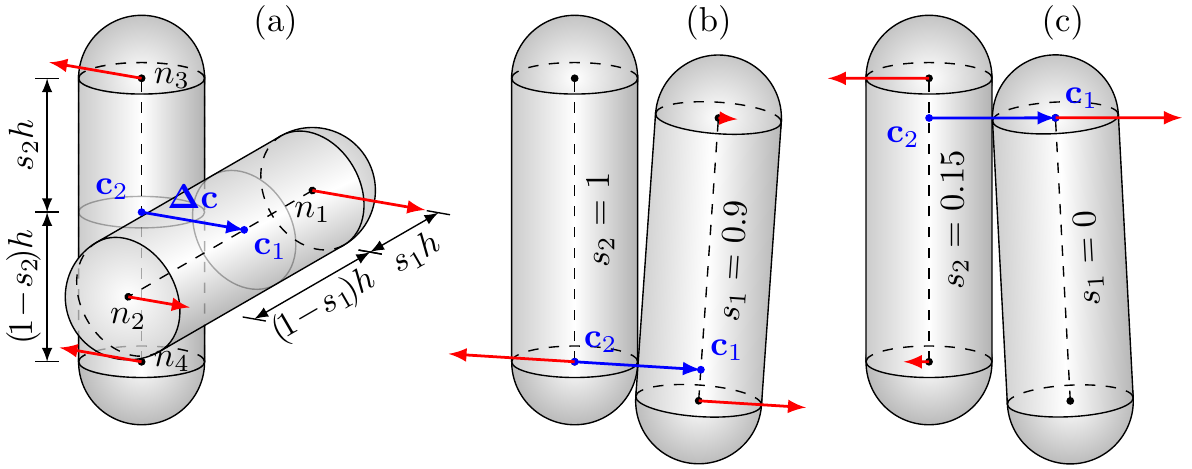}
	  \caption{(a): Wire self-contact, with closest points of approach (blue). The repulsion force (red) is distributed to the four involved nodes. (b)+(c): A small shift in one of the elements can cause the contact parameters $s_1$, $s_2$ to jump, leading to singularity in the force Jacobian.}
	  \label{fig:sc}
	\end{center}
\end{figure*}

Similar to cavity contacts, the spatial self-interaction of the wire is modeled by exchange of Hertzian contact forces that enter the equations of motion (\ref{eq:full_newton}) through the right-hand side ${\bf f}_{\textrm{ext}}$. No moments are transferred, i.e.~all contacts are considered frictionless. The same normal force $f_{\perp}$ as for cavity contacts is used, which is motivated by the observation that most self-contacts occur between almost parallel wire elements. In favor of a compact and general notation, we condense the four involved nodal positions into a single vector
\begin{equation}
{\bf p} = [{\bf p}_{n_1}^{\mathrm{T}},{\bf p}_{n_2}^{\mathrm{T}},{\bf p}_{n_3}^{\mathrm{T}},{\bf p}_{n_4}^{\mathrm{T}}]^{\mathrm{T}}\in\mathbb{R}^{12},
\end{equation}
and apply the same ordering for the resulting force vector ${\bf f}_{\textrm{self},e_1,e_2}\in\mathbb{R}^{12}$ and its Jacobian ${\bf J}_{\textrm{self},e_1,e_2}\in\mathbb{R}^{12\times 12}$ in the following. The Hertzian normal force is distributed to the four nodes by setting
\begin{equation}
\label{eq:fself}
{\bf f}_{\textrm{self},e_1,e_2} = f_{\perp}\left({\bf w}\otimes{\bf n}\right)
\end{equation}
with linearly interpolated nodal weights ${\bf w} = \frac{1}{2}[1-s_1,s_1,-(1-s_2),-s_2]^{\mathrm{T}}$ (see Fig.~\ref{fig:sc}). $\otimes$ denotes the Kronecker product. In other words, the closer a node lies to the contact point between the two elements, the larger the force it experiences, pushing it away from the other element. All possible types of element-element contact are covered by this unified description, be it between coplanar elements, or at any angle. It also inherently handles the cases where the contact point lies on a spherical cap of at least one of the two elements, instead of on their lateral surfaces. Three of the numerous possible contact scenarios are visualized in Fig.~\ref{fig:sc}.

For a Newton-Raphson solver fully accounting for changes in contact forces during each iteration, the derivative of Eq.~(\ref{eq:fself}) with respect to a change of the nodal positions needs to be specified. By the rules of differential calculus, it reads
\begin{align}
\label{eq:jac_self}
{\bf J}_{\textrm{self},e_1,e_2} &= \frac{\partial{\bf f}_{\textrm{self},e_1,e_2}}{\partial{\bf p}} = \left({\bf w}\otimes{\bf n}\right)\frac{\partial f_{\perp}}{\partial{\bf p}} + f_{\perp}\left(\frac{\partial{\bf w}}{\partial{\bf p}}\otimes{\bf n} + {\bf w}\otimes\frac{\partial{\bf n}}{\partial{\bf p}}\right)\in\mathbb{R}^{12\times 12},\\
\frac{\partial f_{\perp}}{\partial{\bf p}} &= -\frac{\partial f_{\perp}}{\partial D}\frac{\partial|{\bf \Delta c}|}{\partial{\bf p}} = -\frac{\pi}{4}E^*h\,{\bf n}^{\mathrm{T}}\frac{\partial{\bf \Delta c}}{\partial{\bf p}}\in\mathbb{R}^{1\times 12},\\
\frac{\partial{\bf n}}{\partial{\bf p}} &= \frac{1}{|{\bf \Delta c}|}\left({\bf I}_3-{\bf n}{\bf n}^{\mathrm{T}}\right)\frac{\partial{\bf \Delta c}}{\partial{\bf p}}\in\mathbb{R}^{3\times 12},\\
\label{eq:ddc_dp}
\frac{\partial{\bf \Delta c}}{\partial{\bf p}} &= \left({\bf w}^{\mathrm{T}}\otimes{\bf I}_3\right) + \big({\bf p}_{n_2} - {\bf p}_{n_1}\big)\frac{\partial s_1}{\partial{\bf p}} - \big({\bf p}_{n_4} - {\bf p}_{n_3}\big)\frac{\partial s_2}{\partial{\bf p}}\in\mathbb{R}^{3\times 12}.
\end{align}
From Eqs.~(\ref{eq:jac_self}\textendash\ref{eq:ddc_dp}), the Jacobian matrix is fully defined, given that $\partial s_1/\partial{\bf p}$ and $\partial s_2/\partial{\bf p}$ can be calculated. However, these are singular if the two wire elements are parallel, since $s_1$ and $s_2$ may jump from one node to the other (see Fig.~\ref{fig:sc}), causing convergence difficulties in the Newton-Raphson method. The exchange of forces between contacting elements is therefore further simplified in the case of implicit integration in time by distributing the normal force equally to all four nodes, i.e.~${\bf w} = \begin{bmatrix}\frac{1}{4},\frac{1}{4},-\frac{1}{4},-\frac{1}{4}\end{bmatrix}^{\mathrm{T}}$, and by discarding $\partial s_1/\partial{\bf p} = \partial s_2/\partial{\bf p} = 0$.

\section{Integration in time}
\label{sec:integration}

\subsection{Newmark methods}

Newmark's family of integration methods \citep{N59} is widely used for solving hyperbolic equilibrium equations in structural dynamics \citep{ZWLX92,R04}. It is particularly well suited in our case because very simple error estimators are available for the predictor-corrector counterparts of the implicit schemes. Although its high precision may not strictly be necessary to account for the dynamic relaxation induced by our choice of lumped masses and damping, it is assisting in efficiently and stably resolving the steep contact forces. For fixed scheme parameters $\beta$ and $\gamma$, the solution vector and its time derivatives are discretized and integrated according to
\begin{align}
\label{eq:newmark_u}
{\bf u}(t+\Delta t) \approx {\bf u}_{t+\Delta t} &= {\bf u}_t + \Delta t\,\dot{{\bf u}}_t + \frac{(\Delta t)^2}{2}\big((1-2\beta)\ddot{{\bf u}}_t + 2\beta\,\ddot{{\bf u}}_{t+\Delta t}\big),\\
\label{eq:newmark_v}
\dot{{\bf u}}(t+\Delta t) \approx \dot{{\bf u}}_{t+\Delta t} &= \dot{{\bf u}}_t + \Delta t\big((1-\gamma)\ddot{{\bf u}}_t + \gamma\,\ddot{{\bf u}}_{t+\Delta t}\big).
\end{align}
$\Delta t$ denotes the finite time step, and the nodal accelerations at time $t$ are given by
\begin{equation}
\ddot{{\bf u}}(t) \approx \ddot{{\bf u}}_t = {\bf M}^{-1}\big({\bf f}_{\textrm{ext}}({\bf u}_t) - {\bf f}_{\textrm{int}}({\bf u}_t) - {\bf C}\dot{{\bf u}}_t\big).
\end{equation}

\subsection{Implicit integration}
\label{sec:impl_int}

In the case that $\beta > 0$, the system of equations of motion (\ref{eq:full_newton}) reduces to the system of nonlinear algebraic equations \citep{R04}
\begin{equation}
\label{eq:nonlinear}
{\bf r}_{t,t+\Delta t} = \overline{{\bf K}}_{t+\Delta t}{\bf u}_{t+\Delta t} - \overline{{\bf f}}_{t,t+\Delta t} = {\bf 0}
\end{equation}
with
\begin{align}
\overline{{\bf K}}_{t+\Delta t} &= a_0\,{\bf M} + a_1\,{\bf C},\\[3pt]
\label{eq:fbar}
\overline{{\bf f}}_{t,t+\Delta t} &= {\bf f}_{\textrm{ext}}({\bf u}_{t+\Delta t}) - {\bf f}_{\textrm{int}}({\bf u}_{t+\Delta t}) + {\bf M}\,{\bf a}_t + {\bf C}\,{\bf b}_t,\\[3pt]
{\bf a}_t &= a_0\,{\bf u}_t + a_2\,\dot{{\bf u}}_t + a_3\,\ddot{{\bf u}}_t,\\[3pt]
\label{eq:auxb}
{\bf b}_t &= a_1\,{\bf u}_t + a_4\,\dot{{\bf u}}_t + a_5\,\ddot{{\bf u}}_t,
\end{align}
and Newmark coefficients
\begin{gather}
a_0 = \frac{1}{\beta\left(\Delta t\right)^2},\quad
a_1 = \frac{\gamma}{\beta\Delta t},\quad
a_2 = \frac{1}{\beta\Delta t},\quad
a_3 = \frac{1}{2\beta}-1,\quad
a_4 = \frac{\gamma}{\beta}-1,\quad
a_5 = \Delta t\left(\frac{\gamma}{2\beta}-1\right).
\end{gather}
In the present implementation, the unconditionally stable \textit{constant-average acceleration method}, that is obtained by setting $\beta=1/4$ and $\gamma=1/2$, is chosen. Our Newton-Raphson (NR) line search algorithm solves Eq.~(\ref{eq:nonlinear}) by taking full account of updated contact forces during the iterations. This requires calculating the residual vector and the Jacobian matrix based on the current iterated guess for ${\bf u}_{t+\Delta t}$. The residual is given by Eq.~(\ref{eq:nonlinear}), where ${\bf f}_{\textrm{ext}}$ is the assembly of nodal contributions from cavity contacts (\ref{eq:fcav}) and self-contacts (\ref{eq:fself}), while ${\bf f}_{\textrm{int}}$ is the assembly of corotated elastic element forces (\ref{eq:cr_fint}). The Jacobian is built according to
\begin{equation}
{\bf J}_{t,t+\Delta t} = {\bf K}_{\textrm{t}} + a_0\,{\bf M} + a_1\,{\bf C} - {\bf J}_{\textrm{ext}},\qquad {\bf J}_{\textrm{ext}} = {\bf J}_{\textrm{cav}} + {\bf J}_{\textrm{self}}
\label{eq:jac}
\end{equation}
in which ${\bf K}_{\textrm{t}}$ is the assembly of element tangent stiffness matrices (\ref{eq:cr_tan_stiff}), ${\bf J}_{\textrm{cav}}$ the assembly of cavity contact Jacobians (\ref{eq:jac_cav}), and ${\bf J}_{\textrm{self}}$ the assembly of self-contact Jacobians (\ref{eq:jac_self}). Upon convergence to the new solution ${\bf u}_{t+\Delta t}$, the corresponding velocity and acceleration vectors are calculated using
\begin{align}
\label{eq:update_ddu}
\ddot{{\bf u}}_{t+\Delta t} &= a_0({\bf u}_{t+\Delta t}-{\bf u}_t) - a_2\,\dot{{\bf u}}_t - a_3\,\ddot{{\bf u}}_t,\\
\label{eq:update_du}
\dot{{\bf u}}_{t+\Delta t} &= \dot{{\bf u}}_t + \Delta t\big((1-\gamma)\ddot{{\bf u}}_t + \gamma\,\ddot{{\bf u}}_{t+\Delta t}\big).
\end{align}
In a simple attempt to optimize performance and to avoid divergence dead ends in energetically difficult configurations, we half the time increment $\Delta t$ as long as more than a predefined number of line search iterations are required for convergence, or double it provided that the number of iterations remains small.

Note that an alternative approach, where contact forces are kept constant during the NR iterations, is conceivable. This would negate the need to compute ${\bf J}_{\textrm{ext}}$, which would just drop out of the formalism, greatly accelerating convergence within a single iteration. See Section \ref{sec:implicit_explicit} for the reason why this is unfeasible in the present context.

\subsection{Explicit integration}
\label{sec:expl_int}

For explicit integration in time, the constant-average acceleration method is implemented in combination with an adaptive time stepping predictor-corrector (APC) algorithm based on an \textit{a posteriori} local error estimator by \cite{ZX91,ZWLX92}. The prediction step is obtained from the explicit part of Eqs.~(\ref{eq:newmark_u}) and (\ref{eq:newmark_v}):
\begin{align}
\label{eq:predict_u}
{\bf u}_{t+\Delta t}^{\textrm{p}} &= {\bf u}_t + \Delta t\,\dot{{\bf u}}_t + \frac{(\Delta t)^2}{2}(1-2\beta)\ddot{{\bf u}}_t,\\
\label{eq:predict_du}
\dot{{\bf u}}_{t+\Delta t}^{\textrm{p}} &= \dot{{\bf u}}_t + \Delta t\,(1-\gamma)\ddot{{\bf u}}_t,\\
\label{eq:predict_ddu}
\ddot{{\bf u}}_{t+\Delta t}^{\textrm{p}} &= {\bf M}^{-1}\big({\bf f}_{\textrm{ext}}({\bf u}_{t+\Delta t}^{\textrm{p}}) - {\bf f}_{\textrm{int}}({\bf u}_{t+\Delta t}^{\textrm{p}}) - {\bf C}\dot{{\bf u}}_{t+\Delta t}^{\textrm{p}}\big).
\end{align}
Owing to lumped masses, ${\bf M}^{-1}(\cdot)$ is a mere component-wise vector multiplication. Based on the predicted acceleration, the displacements and velocities are corrected by the previously omitted part:
\begin{align}
\label{eq:correct_u}
{\bf u}_{t+\Delta t}^{\textrm{c}} &= {\bf u}_{t+\Delta t}^{\textrm{p}} + (\Delta t)^2\beta\,\ddot{{\bf u}}_{t+\Delta t}^{\textrm{p}},\\
\label{eq:correct_du}
\dot{{\bf u}}_{t+\Delta t}^{\textrm{c}} &= \dot{{\bf u}}_{t+\Delta t}^{\textrm{p}} + \Delta t\,\gamma\,\ddot{{\bf u}}_{t+\Delta t}^{\textrm{p}}.
\end{align}
A simple estimator for the relative local error made by performing such a step is then given by 
\begin{equation}
\label{eq:eta}
\eta_{t+\Delta t} = \Big|\beta-\frac{1}{6}\Big|\,\frac{(\Delta t)^2}{u_\textrm{ref}}\,\|\ddot{{\bf u}}_{t+\Delta t}^{\textrm{p}}-\ddot{{\bf u}}_t\|_{\infty},\qquad\beta\ne\frac{1}{6},
\end{equation}
where $\|\cdot\|_{\infty}$ denotes the maximum norm, which is chosen here as a conservative metric. $u_\textrm{ref}$ is a characteristic reference length for normalization, such as the cavity size $u_\textrm{ref}=\overline{R}$ for instance. Ideally, the tradeoff between large time step and small error is dealt with in such a way that the local error is approximately constant over time. Zienkiewicz and Xie found that this can be efficiently achieved by applying the following adaptive time stepping rules, given a desired target value $\overline{\eta}$ for the relative local error, a maximum tolerance $\eta_{\textrm{max}}>\overline{\eta}$, and a lower bound $\eta_{\textrm{min}}<\overline{\eta}$ above which the time step is considered large enough:
\begin{itemize*}
\item If $\eta_{t+\Delta t}\in[\eta_{\textrm{min}},\eta_{\textrm{max}}]$, accept the time step without modifying $\Delta t$.
\item If $\eta_{t+\Delta t} < \eta_{\textrm{min}}$, accept the time step and continue with $\Delta t := \left(\overline{\eta}/\eta_{t+\Delta t}\right)^{1/3}\,\Delta t$.
\item If $\eta_{t+\Delta t} > \eta_{\textrm{max}}$, reject the time step and repeat with $\Delta t := \left(\overline{\eta}/\eta_{t+\Delta t}\right)^{1/3}\,\Delta t$.
\end{itemize*}

\section{Verification and performance}
\label{sec:verification}

The implementation of the finite element program has been realized with the aid of the \texttt{libMesh} \citep{KPSC06} and PETSc \citep{BGMS97} C++ libraries. The applied nonlinear solver is a cubic Newton-Raphson backtracking line search algorithm by \cite{DS96}.

\subsection{Test example}

\begin{figure*}[h]
	\begin{center}
		\includegraphics{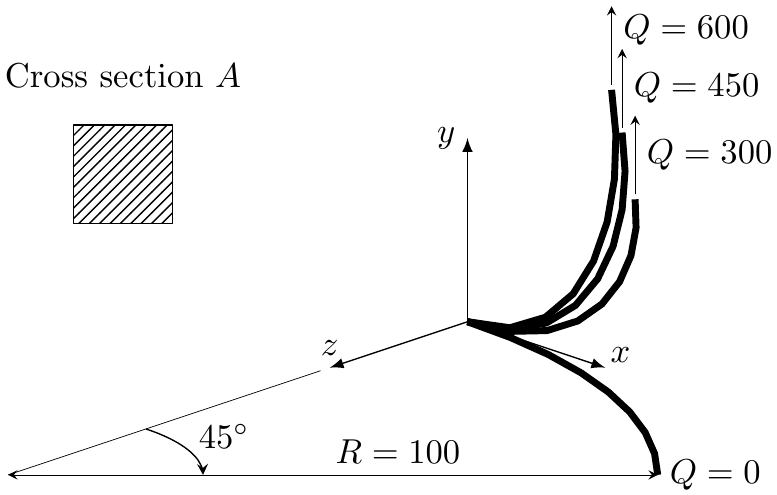}
		\caption{45 degree bend consisting of 8 elements with square cross-section $A$, subject to transverse end loads $Q$.}
		\label{fig:verification}
	\end{center}
\end{figure*}

Verifying and benchmarking nonlinear beam implementations can be done by comparison to various published static test examples \citep[e.g.,][]{BB79}. One of the most prominent ones is reiterated here: The \textit{45 degree bend} scenario is recommended by the National Agency for Finite Element Methods and Standards (NAFEMS, 3DNLG-5) \citep{ABM}. It involves a cantilever beam with square cross-section and non-zero intrinsic rotation variables $|\hat{\theta}_n^0|=h/2R$. The thereby obtained eighth of a circle with radius $R=100$ is subjected to transverse end loads $Q$ leading to a three-dimensional nonlinear response. We have solved it implicitly and fully static, i.e.~with $\rho=c=0$, and depicted it in Fig.~\ref{fig:verification}. The resulting tip positions are listed in Tab.~\ref{tab:tip} for $E=10^7$, $\nu=0$. The reported difference between corotated EBT (CR-EBT) and \cite{C90} appears to be due to the significantly tightened convergence criterion $\|{\bf r}\|_2\le 10^{-8}$ adopted here. Since the order of the theory has almost no effect on such a thin beam, results for thickness 10 are also given, revealing the improved precision of corotated RBT over EBT at large shear stresses.

\newlength{\charw}
\settowidth{\charw}{0}

\begin{table*}[h]
	\centering
	\begin{tabular}{@{}l@{\ \ }c@{\quad}c@{\quad}c@{\ \ }c@{\quad}c@{\quad}c@{\quad}c@{\ \ }c@{\quad}c@{\quad}c@{\quad}c@{}}
	\hline
	\\[-8pt]
	$A=1\!\times\! 1$ & \multicolumn{3}{c}{$Q=300$} & & \multicolumn{3}{c}{$Q=450$} & & \multicolumn{3}{c}{$Q=600$}
	\\[1pt]
	\hline
	\\[-11pt]
	& $p_x$ & $p_y$ & $p_z$ & & $p_x$ & $p_y$ & $p_z$ & & $p_x$ & $p_y$ & $p_z$
	\\[1pt]
	\cline{2-12}
	\\[-8pt]
	CR-RBT (present) & 58.77 & 40.25 & 22.28 & & 52.21 & 48.59 & 18.55 & & 47.11 & 53.58 & 15.73 \\
	CR-EBT (present) & 58.77 & 40.25 & 22.28 & & 52.21 & 48.58 & 18.55 & & 47.11 & 53.57 & 15.73 \\
	\cite{L07} & 58.78 & 40.15 & 22.28 & & 52.24 & 48.46 & 18.56 & & 47.15 & 53.43 & 15.74 \\
	\cite{LW00} & 58.51 & 40.46 & 22.23 & & 51.92 & 48.69 & 18.53 & & 46.82 & 53.6\hspace*{\charw} & 15.76 \\
	\cite{RL98} & 58.58 & 40.31 & 22.16 & & & \textemdash & & & 47.07 & 53.46 & 15.59 \\
	\cite{L92} & 58.8\hspace*{\charw} & 40.1\hspace*{\charw} & 22.3\hspace*{\charw} & & 52.3\hspace*{\charw} & 48.4\hspace*{\charw} & 18.6\hspace*{\charw} & & 47.2\hspace*{\charw} & 53.4\hspace*{\charw} & 15.8\hspace*{\charw} \\
	\cite{C90} & 58.53 & 40.53 & 22.16 & & 51.93 & 48.79 & 18.43 & & 46.84 & 53.71 & 15.61 \\
	\cite{SSD90} & 58.85 & 40.04 & 22.36 & & 52.33 & 48.40 & 18.54 & & 47.27 & 53.34 & 15.88 \\
	\cite{CG88}\hspace*{3pt} & 58.64 & 40.35 & 22.14 & & 52.11 & 48.59 & 18.38 & & 47.04 & 53.50 & 15.55 \\
	\cite{SV86} & 58.84 & 40.08 & 22.33 & & 52.32 & 48.39 & 18.62 & & 47.23 & 53.37 & 15.79 \\
	\cite{BB79} & 59.2\hspace*{\charw} & 39.5\hspace*{\charw} & 22.5\hspace*{\charw} & & & \textemdash & & & 47.2\hspace*{\charw} & 53.4\hspace*{\charw} & 15.9\hspace*{\charw}
	\\[5pt]
	\hline
	\\[-8pt]
	$A=10\!\times\! 10$ & \multicolumn{3}{c}{$Q=3\!\times\! 10^6$} & & \multicolumn{3}{c}{$Q=4.5\!\times\! 10^6$} & & \multicolumn{3}{c}{$Q=6\!\times\! 10^6$}
	\\[1pt]
	\hline
	\\[-8pt]
	CR-RBT (present) & 58.25 & 41.49 & 22.03 & & 51.54 & 49.98 & 18.26 & & 46.35 & 55.09 & 15.43 \\
	CR-EBT (present) & 58.38 & 41.22 & 22.09 & & 51.70 & 49.67 & 18.32 & & 46.54 & 54.75 & 15.48 \\
	\\[-8pt]
	\hline
	\end{tabular}
	\caption{Comparison of tip positions of a clamped 45 degree bend at different end loads and beam thicknesses. Initially ($Q=0$), it is (70.71, 0.0, 29.29).}
	\label{tab:tip}
\end{table*}

\subsection{Implicit vs.~explicit integration}
\label{sec:implicit_explicit}

\begin{figure*}[h]
	\begin{center}
		\includegraphics{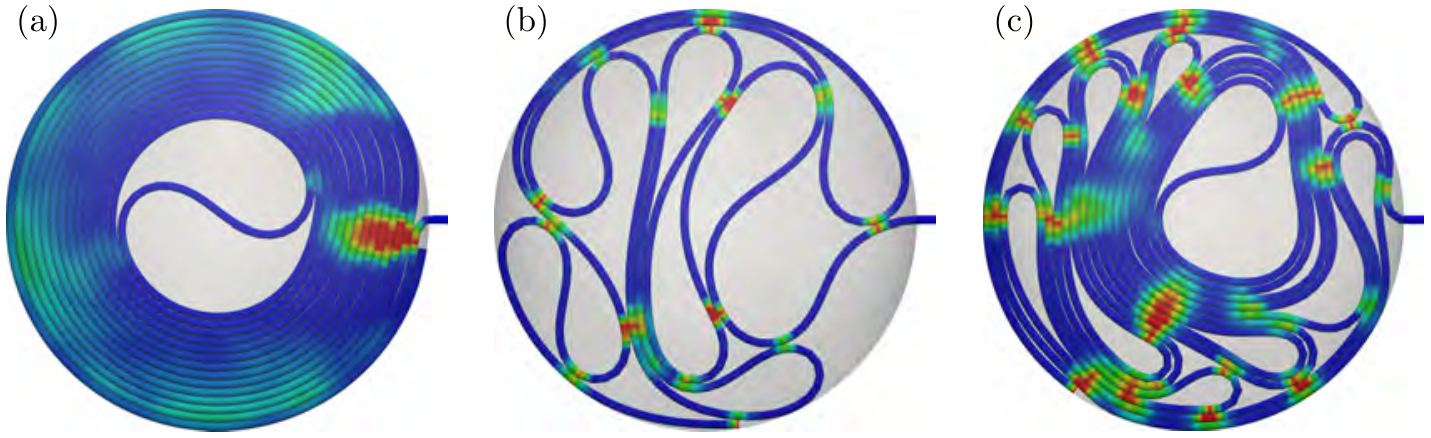}
		\caption{(a, $\phi\approx 0.8$): Low insertion speed, ``spiral'' morphology. Implicit integration converges even at large constant $\Delta t$. (b, $\phi\approx 0.3$): High insertion speed, constant $\Delta t$. Implicit integration convergence failure sets in at very few self-contacts. (c, $\phi\approx 0.6$): High insertion speed, adaptive time stepping. Convergence is retained throughout the simulation. The color encodes the wire-wire contact energy from zero (blue) to high (red). For better visibility, the three energy scales are distinct.}
		\label{fig:impexp}
	\end{center}
\end{figure*}

A general advantage of implicit integration in time over explicit schemes is the ability to cope with large time steps without losing numerical stability. In many problems including the present strong spatial confinement, however, this benefit is compromised by significantly more time-consuming computations for a single time step. The denser a cavity is packed with a wire, the more complex and rough the potential energy landscape gets, with fatal consequences for the convergence behavior of an implicit integration method. As the NR line search solver needs to find a path through energy valleys getting narrower and narrower, there is an increasing chance of starting far enough from the sought energy minimum that no path is found any more. In such situations, adaptive time step control is indispensable for avoiding divergence dead ends and to ensure the found local minimum corresponds to the physical solution. Fig.~\ref{fig:impexp} visualizes the implications in practice. A two-dimensional circular cavity with effective size $R/r=50$ is packed with a frictionless elastic wire, at two different insertion velocities. The ``spiral'' morphology \citep{SWH08} evolving from a slow insertion can be simulated to large packing densities even implicitly at relatively large constant time steps $\Delta t$ (Fig.~\ref{fig:impexp}a). Using the same step size but an insertion speed high enough to introduce dynamic effects from the insertion, the rough valley structure emerging in the local energy landscape inhibits the line search solver from converging as soon as wire-wire contacts reach a certain degree of complexity (Fig.~\ref{fig:impexp}b). Convergence here means meeting one of the convergence criteria within 1000 nonlinear iterations and at line search step sizes $\lambda > \lambda_{\textrm{min}} = 10^{-12}$. The used convergence criteria include $\|{\bf r}\|_2 < 10^{-50}$, $\|{\bf r}\|_2 < 10^{-10}\|{\bf r}_0\|_2$, and $\|{\bf \Delta u}\|_2 < 10^{-10}$, where ${\bf r}_0$ is the initial residual vector, and ${\bf \Delta u}$ is the proposed NR increment. Simulating the ``classic'' morphology \citep{SWH08} to large densities in reasonable CPU time is within the realm of possibility for the fully updated implicit solver only with adaptive time stepping (Fig.~\ref{fig:impexp}c).

Given the steep energy landscape inherently arising from strongly confined wire packings, it may seem tempting to artificially smoothen it by treating contact forces constant during a single NR solver sweep. That is, not updating ${\bf f}_\textrm{ext}$ during solving, and hence setting ${\bf J}_\textrm{ext}={\bf 0}$, as indicated at the end of Section \ref{sec:impl_int}. While this simplifies the formalism and greatly accelerates convergence within each solver sweep, the number of such iterations, that is needed per time step until the residual meets one of the convergence criteria, is large enough to effectively abolish the gain from faster convergence. Reducing this number of iterations to an expedient range requires abridging the time step to unbearably small values. For the sum of these observations, we have desisted from such measures, hence keeping the contact forces fully updated during all solver iterations.

Another inherent, structural reason why an implicit integration in time is in general not feasible is related to the sparsity pattern. As the number of wire-wire contacts approaches $\mathcal{O}(N^2)$ in three-dimensional packings, the global $6(N\!+\!1)\!\times\!6(N\!+\!1)$ system Jacobian will transform from sparse to moderately dense, exposing the iterative NR solver routine to a deteriorating slow-down, assuming that changing contact forces are accounted for.

\begin{figure*}[h]
	\begin{center}
		\includegraphics{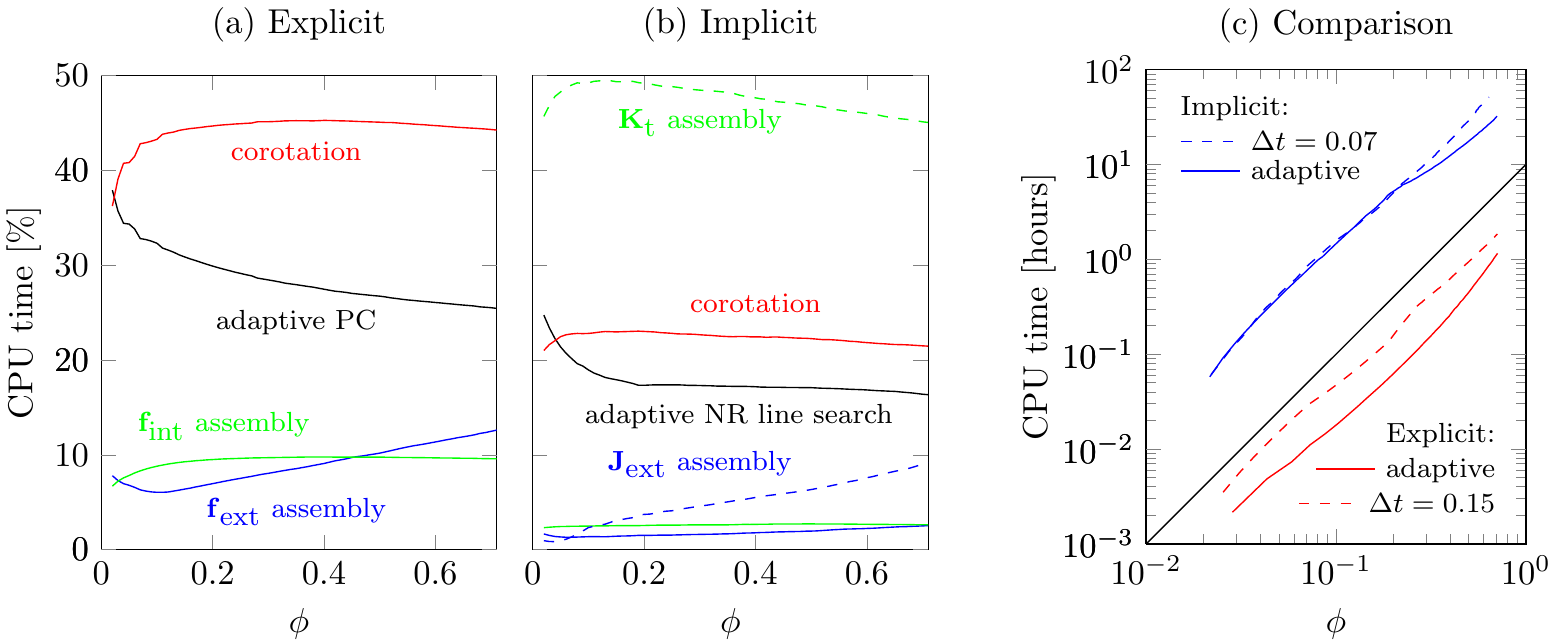}
		\caption{Comparison of computational performance. Left: Integrated relative CPU times for the major program components in the explicit (a) and implicit (b) scheme. See the main text for a detailed description of the shown curves. Right (c): Total integrated CPU times for the explicit (red) and implicit (blue) scheme, with best adaptive time stepping (solid) and best constant time stepping (dashed). The straight solid line is proportional to $\phi^2$.}
		\label{fig:perf}
	\end{center}
\end{figure*}

Fig.~\ref{fig:perf} illustrates how the integration schemes compare in terms of relative and absolute serial performance using a small spherical three-dimensional cavity of effective size $R/r=10$, with $E=10$, $\nu=0.3$, $\rho=1$, $r=1$, $h=2$, $v_\textrm{in}=0.005$, $c=0.1$. The shown curves belong to major disjunct parts of the simulation:
\begin{itemize*}
\item corotation: computation of the corotational transformation matrices ${\bf F}_e$, including $\hat{\bf u}_e$ via Eq.~(\ref{eq:cr_local_rot}), starting from the update of the nodal triads ${\bf T}_n$
\item ${\bf f}_{\textrm{int}}$ assembly: computation of the internal force vector from element contributions (\ref{eq:cr_fint})
\item ${\bf K}_{\textrm{t}}$ assembly: computation of the tangent stiffness matrix from element contributions (\ref{eq:cr_tan_stiff}), via (\ref{eq:cr_stiff})
\item ${\bf f}_{\textrm{ext}}$ assembly: computation of the external force vector from cavity contact forces (\ref{eq:fcav}) and self-contact forces (\ref{eq:fself})
\item ${\bf J}_{\textrm{ext}}$ assembly: computation of the external Jacobian matrix from cavity contact (\ref{eq:jac_cav}) and self-contact (\ref{eq:jac_self})
\item adaptive NR line search: adaptive Newton-Raphson line search integration, following Section \ref{sec:impl_int}
\item adaptive PC: adaptive predictor-corrector integration, following Section \ref{sec:expl_int}
\end{itemize*}
For maximal performance, operations that are common to multiple parts, such as the corotation, are executed only once, and the results are provided to subsequent parts without contributing to their computation time. This allows us to directly compare the two algorithms, since, e.g., the subroutine for the assembly of the corotation matrices is identical. While it eats up most of the computational costs in the explicit APC scheme (Fig.~\ref{fig:perf}a), the implicit method (Fig.~\ref{fig:perf}b) is spending a largely dominant portion of its CPU time assembling the tangent stiffness matrix. As expected, the assembly costs for contact force Jacobians significantly increases with the packing density, while the force vector assembly is almost negligible. The total CPU times are juxtaposed in Fig.~\ref{fig:perf}c. For each scheme, the setup which reaches a density of $\phi=0.7$ the fastest is shown. Numerical stability limits the constant time step to $\Delta t=0.15$ in the explicit case, while the convergence requirements are responsible for the rather small $\Delta t=0.07$ in implicit integration. With adaptivity, larger average time steps can be reached: The fastest settings are $\eta_\textrm{min}=10^{-5}$, $\eta_\textrm{max}=10^{-3}$, resulting in $\Delta t\approx 0.28$ (red solid), and allowing 5 to 40 line search iterations per time step, yielding $\Delta t\approx 0.14$ on average (blue solid). In summary, explicit PC is more than an order of magnitude faster than implicit integration.

\section{Simulation results}
\label{sec:results}

Previously published experiments and simulations of wires subject to spatial confinement share a spherically or cylindrically symmetric container. The goal of the series of simulations presented in the following is to investigate the energy levels arising purely from geometry, and to demonstrate that spherical coils are not naturally assumed in soft confinements. To this end, two different sets of varying ellipsoidal cavities were chosen. The first set consists of isoareal ellipsoids with surface area $S=4000r^2$, at different aspect ratios ($R_x$:$R_y$:$R_z$), imitating an inelastic membrane. The second set of ellipsoids uses the same aspect ratios as the first, but is isochoric at volume $V=2.5\!\times\!10^4\,r^3$ as if elastic cavities were surrounded by an incompressible fluid. The set of cavities comprises a sphere (1:1:1), an oblate spheroid (1:2:2), a prolate spheroid (3:1:1), and a scalene ellipsoid (3:2:1). In all simulations, CR-RBT is used in combination with the explicit adaptive time step integrator. The relevant simulation parameters are $E=10$, $\nu=0.35$, $\rho=1$, $r=1$, $h=3$, $c=0.1$. The wire is inserted at a sufficiently low speed $v_{\textrm{in}}$, allowing the wire to continuously relax to its local energy minimum. Note that even though the insertion is very slow, a fully dynamic treatment of the wire is indispensable, as the large deformation response in the coiled bulk may locally occur must faster than $v_{\textrm{in}}$. This is the case at sudden slip-offs or snap-ins, for instance, where a large amount of elastic energy is suddenly released. 

\begin{figure*}[h]
	\begin{center}
		\includegraphics{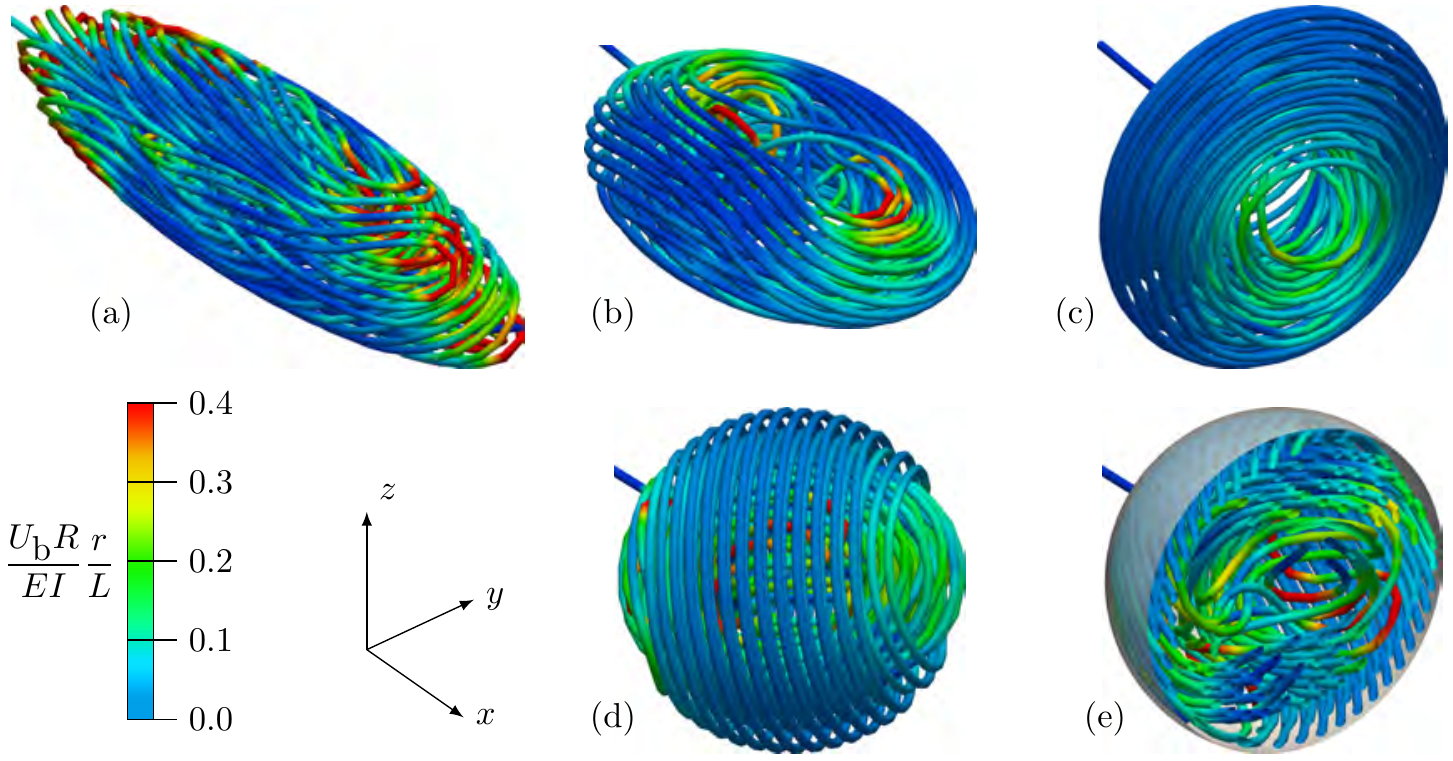}
		\caption{The simulated isochoric ellipsoids with $V=2.5\!\times\!10^4\,r^3$, each at packing density $\phi=0.65$. (a): Prolate (3:1:1). (b): Scalene (3:2:1). (c): Oblate (1:2:2). (d): Sphere (1:1:1). (e): Clipped sphere. The color represents the rescaled bending energy density using $L=\phi V/A$. The wire radius is halved for clearer representation.}
		\label{fig:ellipsoids}
	\end{center}
\end{figure*}

\begin{figure*}[h!]
	\begin{center}
		\includegraphics{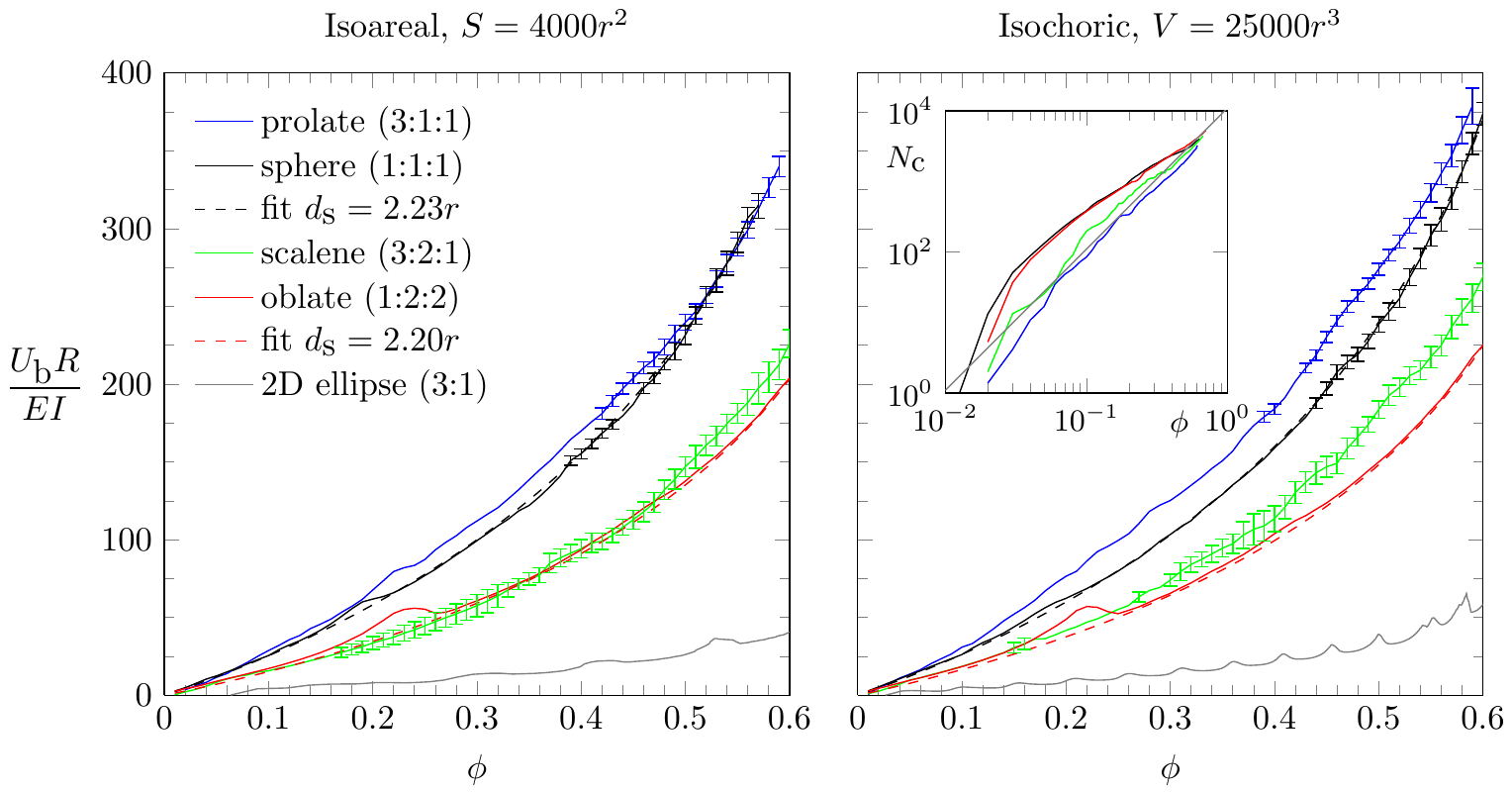}
		\caption{Rescaled bending energy as a function of the packing density for the different ellipsoidal cavities. The inset shows the number of contacts between wire elements. The straight line is proportional to $\phi^2$. Data from 15 runs with moving average filter. Standard errors below 3 are omitted.}
		\label{fig:ubend}
	\end{center}
\end{figure*}

Fig.~\ref{fig:ellipsoids} shows the resulting dense packings for the set of isochoric ellipsoids, and the total bending energies of both sets are juxtaposed in Fig.~\ref{fig:ubend}. The bending energies are rescaled to the dimensionless quantity $U_{\textrm{b}}R/EI$ using the radii of the respective spherical cavities, $R=(3V/4\pi)^{1/3}$. For comparison to the flat scalene limit, the bending energy in two-dimensional ellipses (3:0:1, pseudo-thickness $2r$) of corresponding surface area or ``volume'', respectively, is also plotted. As the contributions from twist and longitudinal compression to the total strain energy are minor compared to the bending energy in all ten cases, $U_{\textrm{b}}$ is safe to be used for distinguishing the total internal energy levels.

Quite naturally, the preferred winding direction imposed by the single short principal axis of the \textit{oblate} confining geometry leads to a highly ordered coil at very low energy (Fig.~\ref{fig:ellipsoids}c). While the \textit{scalene} ellipsoid likewise manages to maintain low energy at early stages due to a preferred winding direction along the path of minimum curvature, emerging figure-eight shapes (Fig.~\ref{fig:ellipsoids}b) then raise the bending energy to a moderately higher level in the isochoric setup. The equally limited spatial extents of the \textit{prolate} cavity perpendicular to its long principal axis, on the other hand, do not dictate any directional preference to the wire other than along the largest axis, leading to complex bulk dynamics at zero friction (Fig.~\ref{fig:ellipsoids}a). The bending energy accommodated by the resulting high-curvature loops near the tips clearly overcompensates the low energy domain near the center, making the prolate spheroid the least favorable of the simulated confinements in terms of strain energy.

Remarkably, the energy contained in a packed \textit{spherical} cavity is comparable to that in a \textit{prolate spheroid}. A cut through the packed sphere (Fig.~\ref{fig:ellipsoids}e) points at strongly disordered bending at high densities inside the shell-like ordered packing as known from \cite{SNWHH11}.

The bending energy in packed spheres or oblate spheroids with $R_x\le R_y=R_z$ can be estimated analytically following the ideas of \cite{PKP03}:
\begin{equation}
\label{eq:ubend_purohit}
U_\textrm{b}(\phi) = \frac{4\pi}{\sqrt{3}}\frac{R_x EI}{d_\textrm{s}(\phi)^2}\left(\log\left[\frac{1+\sqrt{k(\phi)}}{\sqrt{1-k(\phi)}}\right]-\sqrt{k(\phi)}\right)
\end{equation}
where
\begin{equation}
k(\phi) = \left(\frac{3\phi^2}{(2\pi)^2}\left[\frac{d_\textrm{s}(\phi)}{r}\right]^4\right)^{1/3}.
\end{equation}
The segment distance $d_\textrm{s}\ge 2r$, that is in general a function of the packing density, measures the average separation of adjacent wire center lines in cross sections perpendicular to the coiling strands. It can be extracted from simulation snapshots, or calculated \textit{a priori} with a numerical solver in some cases \citep{PKP03}. Motivated by the observation that the density dependence of $d_\textrm{s}$ is very weak \citep{SNWHH11}, we determined it by fitting Eq.~(\ref{eq:ubend_purohit}) to the numerical data in Fig.~\ref{fig:ubend}, assuming constant $d_\textrm{s}$, with excellent agreement in all cases.

\section{Conclusions and outlook}
\label{sec:conclusions}

In the present paper, a detailed description of a finite element model to simulate the packing of elastic wires in two- and three-dimensional cavities has been given. Reddy's locking-free simplified third-order beam theory has been employed to model the bending deflections of the wire, offering an improvement in precision over recent discrete element simulations. Geometric nonlinearity has been introduced to the theory by means of the corotational formulation provided by Crisfield. A quaternion representation has been used to allow for arbitrary wire orientations. Interactions of the wire with hard ellipsoidal cavities and wire-wire contacts were consistently included into the finite element model. By comparison to published numerical results, the correctness of the presented finite element model has been verified.

Two numerical recipes for integrating the hyperbolic equations of motion in time have then been described: The constant-average acceleration method was used as an implicit scheme, and a predictor-corrector method served as the explicit counterpart. Adaptive time stepping is crucial in both methods. For the former, all internal and external forces were supplemented by their Jacobians for the Newton-Raphson iterations, which is a somewhat cumbersome task for self-interacting wire elements. The explicit integration method performs very well, clearly beating implicit integration for the following reasons: The rough potential energy landscape found in densely packed systems with strong body self-interaction calls for relatively small time steps in order to allow fast convergence, and the computational costs accompanying the construction and corotation of the Jacobian impair the overall performance of the implicit scheme.

A few isoareal and isochoric configurations have been simulated. Different ellipsoidal shapes yield different coiling structures of fully elastic, frictionless wires due to asymmetric spatial confinement, the most interesting ones being the scalene and prolate ellipsoids. Published numerical studies concentrated on strict spherical or cylindrical confinement and didn't investigate other geometries. The suboptimal energy state of spherical coils compared to other ellipsoidal shapes, as it was found in this work, hints at a wire preferring non-spherical packing also when inserted into flexible biomaterial antra as they are encountered in real-life applications such as endovascular coiling in the surgical treatment of brain aneurysms, for instance.

Two major physical components determining the morphological characteristics of coiling wires are still missing: (i) A plasticity model is needed to account for realistic elasto-plastic behavior. It should be noted that the present tangent description (\ref{eq:cr_tan_stiff}) is valid only if the internal force may be written as ${\bf f}_{\textrm{int}} = {\bf K}({\bf u})\,{\bf u}$, which will no longer hold. Explicit integration schemes are unaffected by this restriction, however. (ii) The self-contact and cavity contact models also need to be extended by a dry friction model to capture the effects of static and dynamic friction. Intrinsic curvature as a third crucial wire characteristic, on the other hand, is trivial to incorporate into the presented finite elements by setting non-zero local intrinsic angular degrees of freedom in each corotated frame. Together with the suppression of axial rotation at the cavity entrance, pre-curvature controls the amount of order in the wire morphology in packed three-dimensional cavities. Eventually, we expect to find a strong dependence of the coiling morphology on the softness of the cavity, when modeled by a deformable thin shell \citep{VSJWH12}.

\section*{Acknowledgment}

This work was financially supported by the ETH Research Grant ``Packing of Slender Objects in Deformable Confinements'' (ETHIIRA Grant No. ETH-03 10-3).


\begin{thebibliography}{58}
\setlength{\itemsep}{0.2ex}
\expandafter\ifx\csname natexlab\endcsname\relax\def\natexlab#1{#1}\fi
\expandafter\ifx\csname url\endcsname\relax
  \def\url#1{\texttt{#1}}\fi
\expandafter\ifx\csname urlprefix\endcsname\relax\def\urlprefix{URL }\fi
\providecommand{\eprint}[2][]{\url{#2}}
\providecommand{\bibinfo}[2]{#2}
\ifx\xfnm\relax \def\xfnm[#1]{\unskip,\space#1}\fi
\bibitem[{Abaqus()}]{ABM}
Abaqus, \bibinfo{year}{2011}.
\newblock \bibinfo{title}{Abaqus 6.11 {B}enchmarks {M}anual}.
\newblock \bibinfo{organization}{Dassault Syst\`emes Simulia Corp.}.
  \bibinfo{address}{Providence, RI, USA}.
\bibitem[{Balay et~al.(1997)Balay, Gropp, McInnes and Smith}]{BGMS97}
\bibinfo{author}{Balay, S.}, \bibinfo{author}{Gropp, W.D.},
  \bibinfo{author}{McInnes, L.C.}, \bibinfo{author}{Smith, B.F.},
  \bibinfo{year}{1997}.
\newblock \bibinfo{title}{Efficient {M}anagement of {P}arallelism in {O}bject
  {O}riented {N}umerical {S}oftware {L}ibraries}, in: \bibinfo{editor}{Arge,
  E.}, \bibinfo{editor}{Bruaset, A.M.}, \bibinfo{editor}{Langtangen, H.P.}
  (Eds.), \bibinfo{booktitle}{Modern Software Tools for Scientific Computing}.
  \bibinfo{publisher}{Birkh{\"{a}}user Press}, \bibinfo{address}{Cambridge},
  pp. \bibinfo{pages}{163--202}.
\newblock \bibinfo{note}{ISBN 0-8176-3974-8}.
\bibitem[{Bathe and Bolourchi(1979)}]{BB79}
\bibinfo{author}{Bathe, K.J.}, \bibinfo{author}{Bolourchi, S.},
  \bibinfo{year}{1979}.
\newblock \bibinfo{title}{Large displacement analysis of three-dimensional beam
  structures}.
\newblock \bibinfo{journal}{Int. J. Numer. Methods Eng.} \bibinfo{volume}{14},
  \bibinfo{pages}{961--986}.
\bibitem[{Belytschko and Hsieh(1973)}]{BH73}
\bibinfo{author}{Belytschko, T.}, \bibinfo{author}{Hsieh, B.J.},
  \bibinfo{year}{1973}.
\newblock \bibinfo{title}{Non-linear transient finite element analysis with
  convected co-ordinates}.
\newblock \bibinfo{journal}{Int. J. Numer. Methods Eng.} \bibinfo{volume}{7},
  \bibinfo{pages}{255--271}.
\bibitem[{Cardona and Geradin(1988)}]{CG88}
\bibinfo{author}{Cardona, A.}, \bibinfo{author}{Geradin, M.},
  \bibinfo{year}{1988}.
\newblock \bibinfo{title}{A beam finite element non-linear theory with finite
  rotations}.
\newblock \bibinfo{journal}{Int. J. Numer. Methods Eng.} \bibinfo{volume}{26},
  \bibinfo{pages}{2403--2438}.
\bibitem[{Crisfield(1990)}]{C90}
\bibinfo{author}{Crisfield, M.A.}, \bibinfo{year}{1990}.
\newblock \bibinfo{title}{A consistent co-rotational formulation for
  non-linear, three-dimensional, beam-elements}.
\newblock \bibinfo{journal}{Comput. Methods Appl. Mech. Eng.}
  \bibinfo{volume}{81}, \bibinfo{pages}{131--150}.
\bibitem[{Crisfield(1997)}]{C97}
\bibinfo{author}{Crisfield, M.A.}, \bibinfo{year}{1997}.
\newblock \bibinfo{title}{Non-linear Finite Element Analysis of Solids and
  Structures: Volume 2: Advanced Topics}. \bibinfo{publisher}{John Wiley \&
  Sons}, \bibinfo{address}{Chichester}. chapter \bibinfo{chapter}{16\textendash
  17}.
\newblock \bibinfo{note}{ISBN 0-471-95649-X}.
\bibitem[{Dennis and Schnabel(1983)}]{DS96}
\bibinfo{author}{Dennis, Jr., J.E.}, \bibinfo{author}{Schnabel, R.B.},
  \bibinfo{year}{1983}.
\newblock \bibinfo{title}{Numerical Methods for Unconstrained Optimization and
  Nonlinear Equations}. \bibinfo{publisher}{Prentice-Hall},
  \bibinfo{address}{Englewood Cliffs}.
\newblock pp. \bibinfo{pages}{325--327}.
\newblock \bibinfo{note}{ISBN 0-13-627216-9}.
\bibitem[{Donato et~al.(2006)Donato, Oliveira and Gomes}]{DOG06}
\bibinfo{author}{Donato, C.}, \bibinfo{author}{Oliveira, F.},
  \bibinfo{author}{Gomes, M.}, \bibinfo{year}{2006}.
\newblock \bibinfo{title}{Anomalous diffusion on crumpled wires in two
  dimensions}.
\newblock \bibinfo{journal}{Physica A} \bibinfo{volume}{368},
  \bibinfo{pages}{1--6}.
\bibitem[{Donato and Gomes(2007)}]{DG07}
\bibinfo{author}{Donato, C.C.}, \bibinfo{author}{Gomes, M.A.F.},
  \bibinfo{year}{2007}.
\newblock \bibinfo{title}{Condensation of elastic energy in two-dimensional
  packing of wires}.
\newblock \bibinfo{journal}{Phys. Rev. E} \bibinfo{volume}{75},
  \bibinfo{pages}{066113}.
\bibitem[{Donato et~al.(2002)Donato, Gomes and de~Souza}]{DGS02}
\bibinfo{author}{Donato, C.C.}, \bibinfo{author}{Gomes, M.A.F.},
  \bibinfo{author}{de~Souza, R.E.}, \bibinfo{year}{2002}.
\newblock \bibinfo{title}{Crumpled wires in two dimensions}.
\newblock \bibinfo{journal}{Phys. Rev. E} \bibinfo{volume}{66},
  \bibinfo{pages}{015102}.
\bibitem[{Donato et~al.(2003)Donato, Gomes and de~Souza}]{DGS03}
\bibinfo{author}{Donato, C.C.}, \bibinfo{author}{Gomes, M.A.F.},
  \bibinfo{author}{de~Souza, R.E.}, \bibinfo{year}{2003}.
\newblock \bibinfo{title}{Scaling properties in the packing of crumpled wires}.
\newblock \bibinfo{journal}{Phys. Rev. E} \bibinfo{volume}{67},
  \bibinfo{pages}{026110}.
\bibitem[{Eberly(1999)}]{E99}
\bibinfo{author}{Eberly, D.}, \bibinfo{year}{1999}.
\newblock \bibinfo{title}{Distance {B}etween {T}wo {L}ine {S}egments in 3{D}}.
\newblock \bibinfo{note}{\\\url{http://www.geometrictools.com/Documentation/DistanceLine3Line3.pdf}}.
\bibitem[{Eberly(2000)}]{E00}
\bibinfo{author}{Eberly, D.}, \bibinfo{year}{2000}.
\newblock \bibinfo{title}{3D Game Engine Design: A Practical Approach to
  Real-Time Computer Graphics}. \bibinfo{publisher}{Morgan Kaufmann
  Publishers}, \bibinfo{address}{San Francisco}.
\newblock pp. \bibinfo{pages}{43--49}.
\newblock \bibinfo{note}{ISBN 1-55860-593-2}.
\bibitem[{Gomes et~al.(2008a)Gomes, Brito and Ara{\'u}jo}]{GBA08}
\bibinfo{author}{Gomes, M.A.F.}, \bibinfo{author}{Brito, V.P.},
  \bibinfo{author}{Ara{\'u}jo, M.S.}, \bibinfo{year}{2008}a.
\newblock \bibinfo{title}{Geometric properties of crumpled wires and the
  condensed non-solid packing state of very long molecular chains}.
\newblock \bibinfo{journal}{J. Braz. Chem. Soc.} \bibinfo{volume}{19},
  \bibinfo{pages}{293--298}.
\bibitem[{Gomes et~al.(2008b)Gomes, Brito, Coelho and Donato}]{GBCD08}
\bibinfo{author}{Gomes, M.A.F.}, \bibinfo{author}{Brito, V.P.},
  \bibinfo{author}{Coelho, A.S.O.}, \bibinfo{author}{Donato, C.C.},
  \bibinfo{year}{2008}b.
\newblock \bibinfo{title}{Plastic deformation of 2{D} crumpled wires}.
\newblock \bibinfo{journal}{J. Phys. D: Appl. Phys.} \bibinfo{volume}{41},
  \bibinfo{pages}{235408}.
\bibitem[{Grayson and Molineux(2007)}]{GM07}
\bibinfo{author}{Grayson, P.}, \bibinfo{author}{Molineux, I.J.},
  \bibinfo{year}{2007}.
\newblock \bibinfo{title}{Is phage {DNA} `injected' into
  cell\textemdash{}biologists and physicists can agree}.
\newblock \bibinfo{journal}{Curr. Opin. Microbiol.} \bibinfo{volume}{10},
  \bibinfo{pages}{401--409}.
\bibitem[{Guglielmi et~al.(1991a)Guglielmi, Vi{\~n}uela, Dion and
  Duckwiler}]{GVDD91}
\bibinfo{author}{Guglielmi, G.}, \bibinfo{author}{Vi{\~n}uela, F.},
  \bibinfo{author}{Dion, J.}, \bibinfo{author}{Duckwiler, G.},
  \bibinfo{year}{1991}a.
\newblock \bibinfo{title}{Electrothrombosis of saccular aneurysms via
  endovascular approach. {P}art 2: Preliminary clinical experience}.
\newblock \bibinfo{journal}{J.~Neurosurg.} \bibinfo{volume}{75},
  \bibinfo{pages}{8--14}.
\bibitem[{Guglielmi et~al.(1991b)Guglielmi, Vi{\~n}uela, Sepetka and
  Macellari}]{GVSM91}
\bibinfo{author}{Guglielmi, G.}, \bibinfo{author}{Vi{\~n}uela, F.},
  \bibinfo{author}{Sepetka, I.}, \bibinfo{author}{Macellari, V.},
  \bibinfo{year}{1991}b.
\newblock \bibinfo{title}{Electrothrombosis of saccular aneurysms via
  endovascular approach. {P}art 1: Electrochemical basis, technique, and
  experimental results}.
\newblock \bibinfo{journal}{J.~Neurosurg.} \bibinfo{volume}{75},
  \bibinfo{pages}{1--7}.
\bibitem[{Hart(1994)}]{H94}
\bibinfo{author}{Hart, J.C.}, \bibinfo{year}{1994}.
\newblock \bibinfo{title}{Distance to an {E}llipsoid}, in:
  \bibinfo{editor}{Heckbert, P.S.} (Ed.), \bibinfo{booktitle}{Graphics Gems
  IV}. \bibinfo{publisher}{Academic Press}, pp. \bibinfo{pages}{113--119}.
\newblock \bibinfo{note}{ISBN 0-12-336155-9}.
\bibitem[{Johnson(1985)}]{J85}
\bibinfo{author}{Johnson, K.L.}, \bibinfo{year}{1985}.
\newblock \bibinfo{title}{Contact Mechanics}. \bibinfo{publisher}{Cambridge
  University Press}, \bibinfo{address}{Cambridge}.
  chapter~\bibinfo{chapter}{4}.
\newblock \bibinfo{note}{ISBN 0-521-34796-3}.
\bibitem[{Kindt et~al.(2001)Kindt, Tzlil, Ben-Shaul and Gelbart}]{KTBG01}
\bibinfo{author}{Kindt, J.}, \bibinfo{author}{Tzlil, S.},
  \bibinfo{author}{Ben-Shaul, A.}, \bibinfo{author}{Gelbart, W.M.},
  \bibinfo{year}{2001}.
\newblock \bibinfo{title}{{DNA} packaging and ejection forces in
  bacteriophage}.
\newblock \bibinfo{journal}{Proc. Natl. Acad. Sci. U.S.A.}
  \bibinfo{volume}{98}, \bibinfo{pages}{13671--13674}.
\bibitem[{Kirk et~al.(2006)Kirk, Peterson, Stogner and Carey}]{KPSC06}
\bibinfo{author}{Kirk, B.S.}, \bibinfo{author}{Peterson, J.W.},
  \bibinfo{author}{Stogner, R.H.}, \bibinfo{author}{Carey, G.F.},
  \bibinfo{year}{2006}.
\newblock \bibinfo{title}{\texttt{libMesh}: a {C}++ library for parallel
  adaptive mesh refinement/coarsening simulations}.
\newblock \bibinfo{journal}{Eng. Comput.} \bibinfo{volume}{22},
  \bibinfo{pages}{237--254}.
\bibitem[{Leung and Wong(2000)}]{LW00}
\bibinfo{author}{Leung, A.Y.T.}, \bibinfo{author}{Wong, C.K.},
  \bibinfo{year}{2000}.
\newblock \bibinfo{title}{Symmetry {R}eduction of {S}tructures for {L}arge
  {R}otations}.
\newblock \bibinfo{journal}{Adv. Steel Struct.} \bibinfo{volume}{3},
  \bibinfo{pages}{81--102}.
\bibitem[{Li(2007)}]{L07}
\bibinfo{author}{Li, Z.X.}, \bibinfo{year}{2007}.
\newblock \bibinfo{title}{A co-rotational formulation for 3{D} beam element
  using vectorial rotational variables}.
\newblock \bibinfo{journal}{Comput. Mech.} \bibinfo{volume}{39},
  \bibinfo{pages}{309--322}.
\bibitem[{Lo(1992)}]{L92}
\bibinfo{author}{Lo, S.H.}, \bibinfo{year}{1992}.
\newblock \bibinfo{title}{Geometrically nonlinear formulation of 3{D} finite
  strain beam element with large rotations}.
\newblock \bibinfo{journal}{Comput. Struct.} \bibinfo{volume}{44},
  \bibinfo{pages}{147 -- 157}.
\bibitem[{Mattiasson et~al.(1986)Mattiasson, Bengtsson and Samuelsson}]{MBS86}
\bibinfo{author}{Mattiasson, K.}, \bibinfo{author}{Bengtsson, A.},
  \bibinfo{author}{Samuelsson, A.}, \bibinfo{year}{1986}.
\newblock \bibinfo{title}{On the accuracy and efficiency of numerical
  algorithms for geometrically nonlinear structural analysis}, in:
  \bibinfo{editor}{Bergan, P.G.}, \bibinfo{editor}{Bathe, K.J.},
  \bibinfo{editor}{Wunderlich, W.} (Eds.), \bibinfo{booktitle}{Finite Element
  Methods for Nonlinear Problems}. \bibinfo{publisher}{Springer Verlag Berlin},
  \bibinfo{address}{Heidelberg}, pp. \bibinfo{pages}{3--23}.
\newblock \bibinfo{note}{ISBN 3-540-16226-7}.
\bibitem[{Newmark(1959)}]{N59}
\bibinfo{author}{Newmark, N.M.}, \bibinfo{year}{1959}.
\newblock \bibinfo{title}{A {M}ethod of {C}omputation for {S}tructural
  {D}ynamics}.
\newblock \bibinfo{journal}{J. Eng. Mech. Div.} \bibinfo{volume}{85},
  \bibinfo{pages}{67--94}.
\bibitem[{Oran(1973a)}]{C73a}
\bibinfo{author}{Oran, C.}, \bibinfo{year}{1973}a.
\newblock \bibinfo{title}{Tangent {S}tiffness in {P}lane {F}rames}.
\newblock \bibinfo{journal}{J. Struct. Div.} \bibinfo{volume}{99},
  \bibinfo{pages}{973--985}.
\bibitem[{Oran(1973b)}]{C73b}
\bibinfo{author}{Oran, C.}, \bibinfo{year}{1973}b.
\newblock \bibinfo{title}{Tangent {S}tiffness in {S}pace {F}rames}.
\newblock \bibinfo{journal}{J. Struct. Div.} \bibinfo{volume}{99},
  \bibinfo{pages}{987--1001}.
\bibitem[{Petrov and Harvey(2008)}]{PH08}
\bibinfo{author}{Petrov, A.S.}, \bibinfo{author}{Harvey, S.C.},
  \bibinfo{year}{2008}.
\newblock \bibinfo{title}{{P}ackaging {D}ouble-{H}elical {DNA} into {V}iral
  {C}apsids: {S}tructures, {F}orces, and {E}nergetics}.
\newblock \bibinfo{journal}{Biophys. J.} \bibinfo{volume}{95},
  \bibinfo{pages}{497--502}.
\bibitem[{Philipse(1996)}]{P96}
\bibinfo{author}{Philipse, A.P.}, \bibinfo{year}{1996}.
\newblock \bibinfo{title}{The {R}andom {C}ontact {E}quation and {I}ts
  {I}mplications for ({C}olloidal) {R}ods in {P}ackings, {S}uspensions, and
  {A}nisotropic {P}owders}.
\newblock \bibinfo{journal}{Langmuir} \bibinfo{volume}{12},
  \bibinfo{pages}{1127--1133}.
\bibitem[{Prathap and Bhashyam(1982)}]{PB82}
\bibinfo{author}{Prathap, G.}, \bibinfo{author}{Bhashyam, G.R.},
  \bibinfo{year}{1982}.
\newblock \bibinfo{title}{Reduced integration and the shear-flexible beam
  element}.
\newblock \bibinfo{journal}{Int. J. Numer. Methods Eng.} \bibinfo{volume}{18},
  \bibinfo{pages}{195--210}.
\bibitem[{Purohit et~al.(2003)Purohit, Kondev and Phillips}]{PKP03}
\bibinfo{author}{Purohit, P.K.}, \bibinfo{author}{Kondev, J.},
  \bibinfo{author}{Phillips, R.}, \bibinfo{year}{2003}.
\newblock \bibinfo{title}{Mechanics of {DNA} packaging in viruses}.
\newblock \bibinfo{journal}{Proc. Natl. Acad. Sci.} \bibinfo{volume}{100},
  \bibinfo{pages}{3173--3178}.
\bibitem[{Quentrec and Brot(1973)}]{QB73}
\bibinfo{author}{Quentrec, B.}, \bibinfo{author}{Brot, C.},
  \bibinfo{year}{1973}.
\newblock \bibinfo{title}{New {M}ethod for {S}earching for {N}eighbors in
  {M}olecular {D}ynamics {C}omputations}.
\newblock \bibinfo{journal}{J. Comput. Phys.} \bibinfo{volume}{13},
  \bibinfo{pages}{430--432}.
\bibitem[{Rankin and Brogan(1986)}]{RB86}
\bibinfo{author}{Rankin, C.C.}, \bibinfo{author}{Brogan, F.A.},
  \bibinfo{year}{1986}.
\newblock \bibinfo{title}{An {E}lement {I}ndependent {C}orotational {P}rocedure
  for the {T}reatment of {L}arge {R}otations}.
\newblock \bibinfo{journal}{J. Press. Vessel Technol.} \bibinfo{volume}{108},
  \bibinfo{pages}{165--174}.
\bibitem[{Reddy(1997)}]{R97}
\bibinfo{author}{Reddy, J.N.}, \bibinfo{year}{1997}.
\newblock \bibinfo{title}{On locking-free shear deformable beam finite
  elements}.
\newblock \bibinfo{journal}{Comput. Methods Appl. Mech. Eng.}
  \bibinfo{volume}{149}, \bibinfo{pages}{113--132}.
\bibitem[{Reddy(2004)}]{R04}
\bibinfo{author}{Reddy, J.N.}, \bibinfo{year}{2004}.
\newblock \bibinfo{title}{An Introduction to Nonlinear Finite Element
  Analysis}. \bibinfo{publisher}{Oxford University Press},
  \bibinfo{address}{Oxford}.
\newblock pp. \bibinfo{pages}{292--295}.
\newblock \bibinfo{note}{ISBN 0-19-852529-X}.
\bibitem[{Reddy et~al.(1997)Reddy, Wang and Lam}]{RWL97}
\bibinfo{author}{Reddy, J.N.}, \bibinfo{author}{Wang, C.M.},
  \bibinfo{author}{Lam, K.Y.}, \bibinfo{year}{1997}.
\newblock \bibinfo{title}{Unified finite elements based on the classical and
  shear deformation theories of beams and axisymmetric circular plates}.
\newblock \bibinfo{journal}{Commun. Numer. Methods Eng.} \bibinfo{volume}{13},
  \bibinfo{pages}{495--510}.
\bibitem[{Rhim and Lee(1998)}]{RL98}
\bibinfo{author}{Rhim, J.}, \bibinfo{author}{Lee, S.W.}, \bibinfo{year}{1998}.
\newblock \bibinfo{title}{A vectorial approach to computational modelling of
  beams undergoing finite rotations}.
\newblock \bibinfo{journal}{Int. J. Numer. Methods Eng.} \bibinfo{volume}{41},
  \bibinfo{pages}{527--540}.
\bibitem[{Roos et~al.(2007)Roos, Ivanovska, Evilevitch and Wuite}]{RIEW07}
\bibinfo{author}{Roos, W.}, \bibinfo{author}{Ivanovska, I.},
  \bibinfo{author}{Evilevitch, A.}, \bibinfo{author}{Wuite, G.},
  \bibinfo{year}{2007}.
\newblock \bibinfo{title}{Viral capsids: {M}echanical characteristics, genome
  packaging and delivery mechanisms}.
\newblock \bibinfo{journal}{Cell. Mol. Life Sci.} \bibinfo{volume}{64},
  \bibinfo{pages}{1484--1497}.
\bibitem[{Sandhu et~al.(1990)Sandhu, Stevens and Davies}]{SSD90}
\bibinfo{author}{Sandhu, J.S.}, \bibinfo{author}{Stevens, K.A.},
  \bibinfo{author}{Davies, G.}, \bibinfo{year}{1990}.
\newblock \bibinfo{title}{A 3-{D}, co-rotational, curved and twisted beam
  element}.
\newblock \bibinfo{journal}{Comput. Struct.} \bibinfo{volume}{35},
  \bibinfo{pages}{69 -- 79}.
\bibitem[{Shabana(2008)}]{S08}
\bibinfo{author}{Shabana, A.A.}, \bibinfo{year}{2008}.
\newblock \bibinfo{title}{Computational Continuum Mechanics}.
  \bibinfo{publisher}{Cambridge University Press},
  \bibinfo{address}{Cambridge}.
\newblock pp. \bibinfo{pages}{118--120}.
\newblock \bibinfo{note}{ISBN 0-521-88569-8}.
\bibitem[{Simo and Vu-Quoc(1986)}]{SV86}
\bibinfo{author}{Simo, J.C.}, \bibinfo{author}{Vu-Quoc, L.},
  \bibinfo{year}{1986}.
\newblock \bibinfo{title}{A three-dimensional finite-strain rod model. {P}art
  {II}: {C}omputational aspects}.
\newblock \bibinfo{journal}{Comput. Methods Appl. Mech. Eng.}
  \bibinfo{volume}{58}, \bibinfo{pages}{79--116}.
\bibitem[{Stoop et~al.(2011)Stoop, Najafi, Wittel, Habibi and
  Herrmann}]{SNWHH11}
\bibinfo{author}{Stoop, N.}, \bibinfo{author}{Najafi, J.},
  \bibinfo{author}{Wittel, F.K.}, \bibinfo{author}{Habibi, M.},
  \bibinfo{author}{Herrmann, H.J.}, \bibinfo{year}{2011}.
\newblock \bibinfo{title}{Packing of {E}lastic {W}ires in {S}pherical
  {C}avities}.
\newblock \bibinfo{journal}{Phys. Rev. Lett.} \bibinfo{volume}{106},
  \bibinfo{pages}{214102}.
\bibitem[{Stoop et~al.(2008)Stoop, Wittel and Herrmann}]{SWH08}
\bibinfo{author}{Stoop, N.}, \bibinfo{author}{Wittel, F.K.},
  \bibinfo{author}{Herrmann, H.J.}, \bibinfo{year}{2008}.
\newblock \bibinfo{title}{Morphological {P}hases of {C}rumpled {W}ire}.
\newblock \bibinfo{journal}{Phys. Rev. Lett.} \bibinfo{volume}{101},
  \bibinfo{pages}{094101}.
\bibitem[{Sunday(2001)}]{S01}
\bibinfo{author}{Sunday, D.}, \bibinfo{year}{2001}.
\newblock \bibinfo{title}{Distance between {L}ines and {S}egments with their
  {C}losest {P}oint of {A}pproach}.
\newblock \bibinfo{note}{\\\url{http://www.softsurfer.com/Archive/algorithm_0106/algorithm_0106.htm}}.
\bibitem[{Tamatani et~al.(2002)Tamatani, Ito, Abe, Koike, Takeuchi and
  Tanaka}]{TIAKTT02}
\bibinfo{author}{Tamatani, S.}, \bibinfo{author}{Ito, Y.},
  \bibinfo{author}{Abe, H.}, \bibinfo{author}{Koike, T.},
  \bibinfo{author}{Takeuchi, S.}, \bibinfo{author}{Tanaka, R.},
  \bibinfo{year}{2002}.
\newblock \bibinfo{title}{Evaluation of the {S}tability of {A}neurysms after
  {E}mbolization {U}sing {D}etachable {C}oils: {C}orrelation between
  {S}tability of {A}neurysms and {E}mbolized {V}olume of {A}neurysms}.
\newblock \bibinfo{journal}{AJNR Am.~J.~Neuroradiol.} \bibinfo{volume}{23},
  \bibinfo{pages}{762--767}.
\bibitem[{Teh and Clarke(1998)}]{TC98}
\bibinfo{author}{Teh, L.H.}, \bibinfo{author}{Clarke, M.J.},
  \bibinfo{year}{1998}.
\newblock \bibinfo{title}{Co-rotational and {L}agrangian formulations for
  elastic three-dimensional beam finite elements}.
\newblock \bibinfo{journal}{J. Construct. Steel Res.} \bibinfo{volume}{48},
  \bibinfo{pages}{123--144}.
\bibitem[{Tessler and Dong(1981)}]{TD81}
\bibinfo{author}{Tessler, A.}, \bibinfo{author}{Dong, S.},
  \bibinfo{year}{1981}.
\newblock \bibinfo{title}{On a hierarchy of conforming {T}imoshenko beam
  elements}.
\newblock \bibinfo{journal}{Comput. Struct.} \bibinfo{volume}{14},
  \bibinfo{pages}{335--344}.
\bibitem[{Timoshenko(1921)}]{T21}
\bibinfo{author}{Timoshenko, S.P.}, \bibinfo{year}{1921}.
\newblock \bibinfo{title}{On the correction for shear of the differential
  equation for transverse vibrations of prismatic bars}.
\newblock \bibinfo{journal}{Philos. Mag.} \bibinfo{volume}{41},
  \bibinfo{pages}{744--746}.
\bibitem[{Timoshenko(1922)}]{T22}
\bibinfo{author}{Timoshenko, S.P.}, \bibinfo{year}{1922}.
\newblock \bibinfo{title}{On the transverse vibrations of bars of uniform
  cross-section}.
\newblock \bibinfo{journal}{Philos. Mag.} \bibinfo{volume}{43},
  \bibinfo{pages}{125--131}.
\bibitem[{Tong et~al.(1971)Tong, Pian and Bucciarblli}]{TPB71}
\bibinfo{author}{Tong, P.}, \bibinfo{author}{Pian, T.H.H.},
  \bibinfo{author}{Bucciarblli, L.L.}, \bibinfo{year}{1971}.
\newblock \bibinfo{title}{Mode shapes and frequencies by finite element method
  using consistent and lumped masses}.
\newblock \bibinfo{journal}{Comput. Struct.} \bibinfo{volume}{1},
  \bibinfo{pages}{623--638}.
\bibitem[{Urthaler and Reddy(2005)}]{UR05}
\bibinfo{author}{Urthaler, Y.}, \bibinfo{author}{Reddy, J.N.},
  \bibinfo{year}{2005}.
\newblock \bibinfo{title}{A corotational finite element formulation for the
  analysis of planar beams}.
\newblock \bibinfo{journal}{Commun. Numer. Methods Eng.} \bibinfo{volume}{21},
  \bibinfo{pages}{553--570}.
\bibitem[{Vetter et~al.(2012)Vetter, Stoop, Jenni, Wittel and
  Herrmann}]{VSJWH12}
\bibinfo{author}{Vetter, R.}, \bibinfo{author}{Stoop, N.},
  \bibinfo{author}{Jenni, T.}, \bibinfo{author}{Wittel, F.K.},
  \bibinfo{author}{Herrmann, H.J.}, \bibinfo{year}{2012}.
\newblock \bibinfo{title}{Subdivision shell elements with anisotropic growth}.
\newblock \bibinfo{note}{Manuscript in preparation}.
\bibitem[{Wempner(1969)}]{W69}
\bibinfo{author}{Wempner, G.}, \bibinfo{year}{1969}.
\newblock \bibinfo{title}{Finite elements, finite rotations and small strains
  of flexible shells}.
\newblock \bibinfo{journal}{Int. J. Solids Struct.} \bibinfo{volume}{5},
  \bibinfo{pages}{117--153}.
\bibitem[{Zeng et~al.(1992)Zeng, Wiberg, Li and Xie}]{ZWLX92}
\bibinfo{author}{Zeng, L.F.}, \bibinfo{author}{Wiberg, N.E.},
  \bibinfo{author}{Li, X.D.}, \bibinfo{author}{Xie, Y.M.},
  \bibinfo{year}{1992}.
\newblock \bibinfo{title}{\textit{A posteriori} local error estimation and
  adaptive time-stepping for {N}ewmark integration in dynamic analysis}.
\newblock \bibinfo{journal}{Earthquake Engng Struct. Dyn.}
  \bibinfo{volume}{21}, \bibinfo{pages}{555--571}.
\bibitem[{Zienkiewicz and Xie(1991)}]{ZX91}
\bibinfo{author}{Zienkiewicz, O.C.}, \bibinfo{author}{Xie, Y.M.},
  \bibinfo{year}{1991}.
\newblock \bibinfo{title}{A simple error estimator and adaptive time stepping
  procedure for dynamic analysis}.
\newblock \bibinfo{journal}{Earthquake Engng Struct. Dyn.}
  \bibinfo{volume}{20}, \bibinfo{pages}{871--887}.

\end{thebibliography}
\end{document}